\documentclass{aa}

\usepackage{graphicx}
\usepackage{txfonts}
\usepackage{natbib}
\usepackage{subfig}
\usepackage{siunitx}
\usepackage{adjustbox}

\newcommand{\angstrom}{\mbox{\normalfont\AA}~}
\newcommand{\aminvsg}{$a_{\mathrm{min,\,a-C}}$}
\newcommand{\abvsg}{M$_{\mathrm{a-C}}$/M$_{\mathrm{H}}$}
\newcommand{\lpdr}{$l_{\rm{PDR}}$}

\newcommand{\mum}{$\mu$m}
\newcommand{\cm}{a-C}

\newcommand{\CM}{a-C:H/a-C}
\newcommand{\py}{a-Sil/a-C}

\newcommand{\G}{$G_{0}$}

\newcommand{\chimin}{$\chi^{2}_{\mathrm{min}}$}
\newcommand{\chimind}{$\chi^{2}_{\mathrm{min,\,2D}}$}
\newcommand{\chimindab}{$\chi^{2}_{\mathrm{min,\,2D}}\left(M_{\mathrm{a-C}}/M_{\mathrm{H}}\right)$}
\newcommand{\chii}{$\chi^{2}$}

\newcommand{\MIPSun}{$\mathrm{MIPS}_{24}$}
\newcommand{\MIPSdeux}{$\mathrm{PACS}_{70}$}
\newcommand{\MIPStrois}{$\mathrm{PACS}_{160}$}
\newcommand{\SPIREun}{$\mathrm{SPIRE}_{250}$}
\newcommand{\SPIREdeux}{$\mathrm{SPIRE}_{350}$}
\newcommand{\SPIREtrois}{$\mathrm{SPIRE}_{500}$}
\newcommand{\IRACun}{$\mathrm{IRAC}_{3.6}$}
\newcommand{\IRACdeux}{$\mathrm{IRAC}_{4.5}$}
\newcommand{\IRACtrois}{$\mathrm{IRAC}_{5.8}$}
\newcommand{\IRACquatre}{$\mathrm{IRAC}_{8.0}$}

\newcommand{\herschel}{\textit{Herschel}}
\newcommand{\spitzer}{\textit{Spitzer}}

\begin{document}

   \title{Dust evolution across the Horsehead Nebula}

   \subtitle{}

   \author{T. Schirmer.
          \inst{1}
          \and
          A. Abergel\inst{1}
          \and 
          L. Verstraete\inst{1}
          \and
          N. Ysard\inst{1}
          \and 
          M. Juvela\inst{2}
          \and 
          A. P. Jones\inst{1}
          \and
          E. Habart\inst{1}
          }

   \institute{Université Paris-Saclay, CNRS,  Institut d'astrophysique spatiale, 91405, Orsay, France \\
   \email{thiebaut.schirmer@ias.u-psud.fr}
   \and
   Department of Physics, PO Box 64, University of Helsinki, 00014 Helsinki, Finland
             }

   \date{Received 12 March 2020; accepted ??}

 
  \abstract
   {Micro-physical processes on interstellar dust surfaces are tightly
   connected to dust properties (i.e. dust composition, size and shape) and play a key 
   role in numerous phenomena in the interstellar medium (ISM). The large disparity in
   physical conditions (i.e. density, gas temperature) in the ISM triggers an evolution
   of dust properties. The analysis of how dust evolves with the physical conditions is
   a stepping-stone towards a more thorough understanding of interstellar dust.}
   {The aim
   of this paper is to highlight dust evolution in the Horsehead Nebula PDR region.}
   {We use \textit{Spitzer}/IRAC (3.6, 
   4.5, 5.8 and 8 \mum), \textit{Spitzer}/MIPS (24 \mum) together with \textit{Herschel}/PACS
   (70 and 160 \mum) and \textit{Herschel}/SPIRE (250, 350 and 500 \mum) to map the
   spatial distribution of dust in the Horsehead over the entire emission spectral range. We model dust emission and
   scattering using the THEMIS interstellar dust model together with the 3D radiative
   transfer code SOC.}
   {We find that the nano-grains dust-to-gas ratio in the irradiated outer part of the Horsehead is 6 to 10 times lower than in the diffuse ISM. Their minimum size is 2 to 2.25 times larger than in the diffuse ISM and the power-law exponent of their size distribution, 1.1 to 1.4 times lower than in the diffuse ISM. Regarding the denser part of the Horsehead,
   it is necessary to use evolved grains (i.e. aggregates, with or without an ice mantle).}
   {It is not possible to explain the observations using grains from the diffuse medium. We therefore propose the following scenario to explain our results. In the outer part of the Horsehead, all the nano-grains have not yet had time to re-form completely through 
   photo-fragmentation of aggregates and the smallest of the nano-grains that 
   are sensitive to the radiation field are photo-destroyed. In the inner part of the Horsehead, grains most likely consist of multi-compositional, mantled aggregates. }


   \maketitle
%

\section{Introduction}

Interstellar dust plays an essential role within the interstellar
medium (ISM) through different microphysical processes happening on dust
surfaces that can heat the gas, such as the photoelectric effect  \citep[e.g.][]{bakes_photoelectric_1994,
weingartner_photoelectric_2001}, or cool the gas through gas-grain collisions 
\citep{burke_gas-grain_1983}. By acting as a catalyst, allowing atoms and molecules to react on its surface, dust is strongly involved in the 
chemistry of the ISM
\citep[e.g.][]{hollenbach_surface_1971,bron_surface_2014, jones_h_2015, wakelam_h$_2$_2017}. Also, dust plays a role in the 
redistribution of
UV-visible stellar radiation into IR-mm radiation, a process that depends on the dust mass
and the volume of dust grains
\citep[e.g.][]{draine_interstellar_2003,compiegne_global_2011}. The efficiency of these
processes strongly depends on the dust properties, such as the grain size,
composition and shape. It is therefore important to constrain dust 
properties in order to understand the different phenomena that take
place in the ISM. To this purpose, several dust models have been developed and are
classified in three
categories. Those composed of silicate and graphite \citep[e.g][]{mathis_size_1977, draine_optical_1984, kim_size_1994} with an 
extension of these models, using PAHs (polycyclic aromatic hydrocarbons) \citep[e.g.][]{siebenmorgen_dust_1992, li_ultrasmall_2001,weingartner_dust_2001}. As a result of fragmentation and coagulation
processes in the ISM, dust models with grains that
have a core-mantle structure \citep[e.g.][]{desert_interstellar_1990,jones_structure_1990,li_unified_1997} and 
dust composite models composed of silicate and carbon grain aggregates 
\citep[e.g.][]{mathis_composite_1989, zubko_interstellar_2004} have been proposed. In
this paper, we use the THEMIS dust model \citep[see][]{jones_evolution_2013,kohler_hidden_2014,jones_cycling_2014,kohler_dust_2015,ysard_dust_2015,jones_global_2017}, developed in combination with the results of laboratory experiments and astronomical observations. The
cornerstone of this model is its self-consistent view of the evolution of the dust constituents through the ISM. This view is required for 
understanding dust evolution in response to the local ISM conditions 
(i.e. density, radiation field).

Some of the first evidence of dust evolution was shown by \cite{fitzpatrick_analysis_1986} through the variation in the 2175 \angstrom
interstellar bump from diffuse ($R_V=3.1$) to denser regions (up to $R_V$ $\sim$ 5.5). Similarly, other studies \citep[e.g.][]{cardelli_relationship_1989, cardelli_absolute_1991, campeggio_total_2007} found the same variations
that have for the first time been explained by
\cite{kim_size_1994} by stating that these observations are consistent with
a decrease in the carbonaceous nano-grains abundance (relative to the gas) together
with an increase in larger grain abundance. It is also possible to follow dust evolution from its emission in the mid-IR (due to stochastically heated nano-grains) and in the far-IR (where large grains at thermal equilibrium emit). This has led to a wealth of studies \citep[e.g.][]{boulanger_variations_1990, laureijs_iras_1991, abergel_comparative_1994, bernard_pronaos_1999, stepnik_evolution_2003, flagey_evidence_2009} revealing that nano-grains disappear in dense regions as they
coagulate onto larger grains. Dust evolution is also highlighted by variation of its far-IR opacity with the local environment \citep[e.g.][]{juvela_galactic_2011,planck_collaboration_planck_2011-1,martin_evidence_2012,roy_changes_2013,ysard_variation_2013,kohler_dust_2015,juvela_galactic_2015} explained with dust coagulation and accretion of ice mantles, a scenario which is 
supported by numerical simulations of dust evolution in dense regions
\citep[e.g.][]{ossenkopf_dust_1994,ormel_dust_2011, kohler_dust_2015}.

Photon-dominated regions (PDRs) \citep{hollenbach_dense_1997, hollenbach_photodissociation_1999} correspond to the interface
between HII regions and molecular clouds that are irradiated
by energetic stars close by. In these regions, the physical conditions are strongly
constrated hence PDRs are a unique place to study how do dust, gas
and local physical conditions evolve with depth. Based on dust
emission variations in the mid-IR observed with \spitzer~in several PDRs (Ced 201, NGC 7023, $\rho$ Ophiuchi West filament), \cite{berne_analysis_2007} construed that such variations can be explained by the photo-processing of carbonaceous nano-grains, a scenario later reinforced in other PDRs
\citep{abergel_evolution_2010, pilleri_evaporating_2012, boersma_properties_2014, pilleri_variations_2015}. Using far-IR observations from \herschel, together with the near and mid-IR 
observations from \spitzer, \cite{arab_evolution_2012} found that the carbonaceous nano-grains abundance decreases together with an increase in the opacity of the large grains in
the Orion bar. They claimed that these variations are likely due to coagulation
processes in the denser part of this region. Evidence of dust evolution has also been shown in IC 63 based on 
extinction mapping \citep{van_de_putte_evidence_2019}. 

In this paper, we focus on a well-known PDR, the Horsehead, that has previously been studied from the 
perspective of dust observations
\citep[e.g.][]{abergel_isocam_2003,teyssier_carbon_2004, compiegne_aromatic_2007, pety_are_2005, compiegne_dust_2008,arab_evolution_2012-2}, gas observations \citep[e.g.][]{habart_density_2005, goicoechea_low_2006,gerin_hco_2009,guzman_h2co_2011, pety_iram-30_2012,ohashi_mapping_2013,le_gal_new_2017} and laboratory experiments \citep{alata_vacuum_2015}. The most important question we try to answer is how the dust properties change with physical conditions. Thus, is it possible to understand these observations with grains from the diffuse ISM? Otherwise, is there a viable dust evolution scenario that can explain the observations and is consistent with the physical conditions in the Horsehead?

The paper is organised as follows. In Sect.\,\ref{sec:PDR}, we describe
the previous studies and the observations of the Horsehead. In Sect.\,\ref{sec:models_tools}, we detail the dust model we use, THEMIS, 
as well as the local dust emission tool, DustEM. We also present the effects
of variations in dust properties on its emission
in the optically thin case with DustEM in order to disentangle variations
in the dust spectrum due to changes in dust properties and those due to
radiative transfer effects. In Sect.\,\ref{sec:dust_emission_radiative_transfer},
we present SOC, the 3D radiative transfer code we use, as well as the effect of
variations in the dust parameters on dust emission in the optically thick
case, in the case of the Horsehead. In
Sect.\,\ref{sec:comparison_observations}, we compare our model with the
observations and present the best parameters 
we obtain. In Sect.\,\ref{sec:discussion}, we discuss our results and propose a scenario of dust evolution in the Horsehead. Finally, we
present in Sect.\,\ref{sec:conclusion} our conclusions. 

\section{A prototypical PDR: the Horsehead}
\label{sec:PDR}
As physical conditions are strongly constrasted and spatially resolved
in nearby photodominated regions, they are the ideal place to study 
dust evolution as a function of physical conditions. First, we introduce
the different studies that have been made of the Horsehead;
second, we present the observations of the Horsehead obtained with 
\spitzer~and \herschel; third, we describe the density profile that we
use to perform radiative transfer across the Horsehead.

\subsection{A well studied PDR}
The Horsehead is an archetypal PDR situated at $\sim$ 400 pc \citep{anthony-twarog_h-beta_1982} that is illuminated by the 
binary star $\sigma$-Orionis which is an O9.5V binary system
\citep{warren_photometric_1977} with an effective temperature of 
$T_{\mathrm{eff}}\sim$~34600~K 
\citep{schaerer_combined_1997} located at a projected distance  $d_{\mathrm{edge}}\sim$~3.5~pc from the Horsehead
edge. Observations of the Horsehead have been made in the visible \citep[e.g.][]{de_boer_diffuse_1983, neckel_spectroscopic_1985} and at millimeter
wavelengths for $^{12}$CO and $^{13}$CO \citep[e.g.][]{milman_co_1973}, $^{12}$CO \citep[e.g.][]{stark_co_1982}, NH$_{3}$ \citep[e.g.][]{sandell_young_1986}, CS
\citep[e.g.][]{lada_unbiased_1991}, C$^{+}$ \citep[e.g.][]{zhou_[c_1993} and $^{13}$CO
\citep[e.g.][]{kramer_structure_1996}.

Later, mid-IR observations \citep{abergel_isocam_2003} with ISOCAM highlighted that the 
Horsehead is likely to be seen edge-on hence offers us a unique opportunity 
to study dust, gas and the evolution of local physical conditions with depth
into the Horsehead.
This has led to many studies at millimeter wavelengths
for CO \citep{pound_looking_2003}, C$^{18}$O \citep{hily-blant_velocity_2005},
CS, C$^{34}$S and HCS$^{+}$ \citep{goicoechea_low_2006}, CI and CO
\citep{philipp_submillimeter_2006}, DCO$^{+}$ \citep{pety_deuterium_2007},
HCO and H$^{13}$CO$^{+}$ \citep{gerin_hco_2009}, H$^{13}$CO$^{+}$, DCO$^{+}$ and
HCO$^{+}$ \citep{goicoechea_ionization_2009}, H$_{2}$CO \citep{guzman_h2co_2011},
CF$^{+}$ \citep{guzman_iram-30m_2012}, l-C$_{3}$H$^{+}$ \citep{pety_iram-30_2012},
CH$_{3}$CN, HC$_{3}$N and C$_{3}$N \citep{gratier_iram-30_2013}, H$_{2}$CO and 
CH$_{3}$OH \citep{guzman_iram-30_2013}, NH$_{3}$ \citep{ohashi_mapping_2013},
HCOOH, CH$_{2}$CO, CH$_{3}$CHO and CH$_{3}$CCH \citep{le_gal_new_2017}.

Regarding dust, \cite{teyssier_carbon_2004} found that although small hydrocarbons
are supposed 
to be photo-destroyed by the intense UV field at the edge of the Horsehead, they are
still existing. They suggest that the photo-erosion of carbonaceous nano-grains into small
hydrocarbons is more efficient than the photo-destruction of small hydrocarbons at
the Horsehead edge. This scenario is reinforced by observations in \cite{pety_are_2005} as
they found hydrocarbons such as CCH, c-C$_{3}$H$_{2}$ and C$_{4}$H in the UV-irradiated 
outer part of the Horsehead. It is also supported by laboratory experiments on thermal processed and UV-irradiated 
dust grains analogues \citep[see][]{smith_optical_1984, zubko_interstellar_2004, alata_vacuum_2014,alata_vacuum_2015,duley_small_2015}. Based on
Spitzer observations, \cite{compiegne_aromatic_2007}
proposed a scenario where PAHs survive in HII regions and 
\cite{compiegne_dust_2008} construed that spectral variations in the mid-IR cannot
only be explained by radiative transfer effects and therefore are a consequence 
of dust evolution across the Horsehead. 

\begin{figure}[h]
\centering
	\includegraphics[width=0.5\textwidth, trim={0 0cm 0cm 0cm},clip]{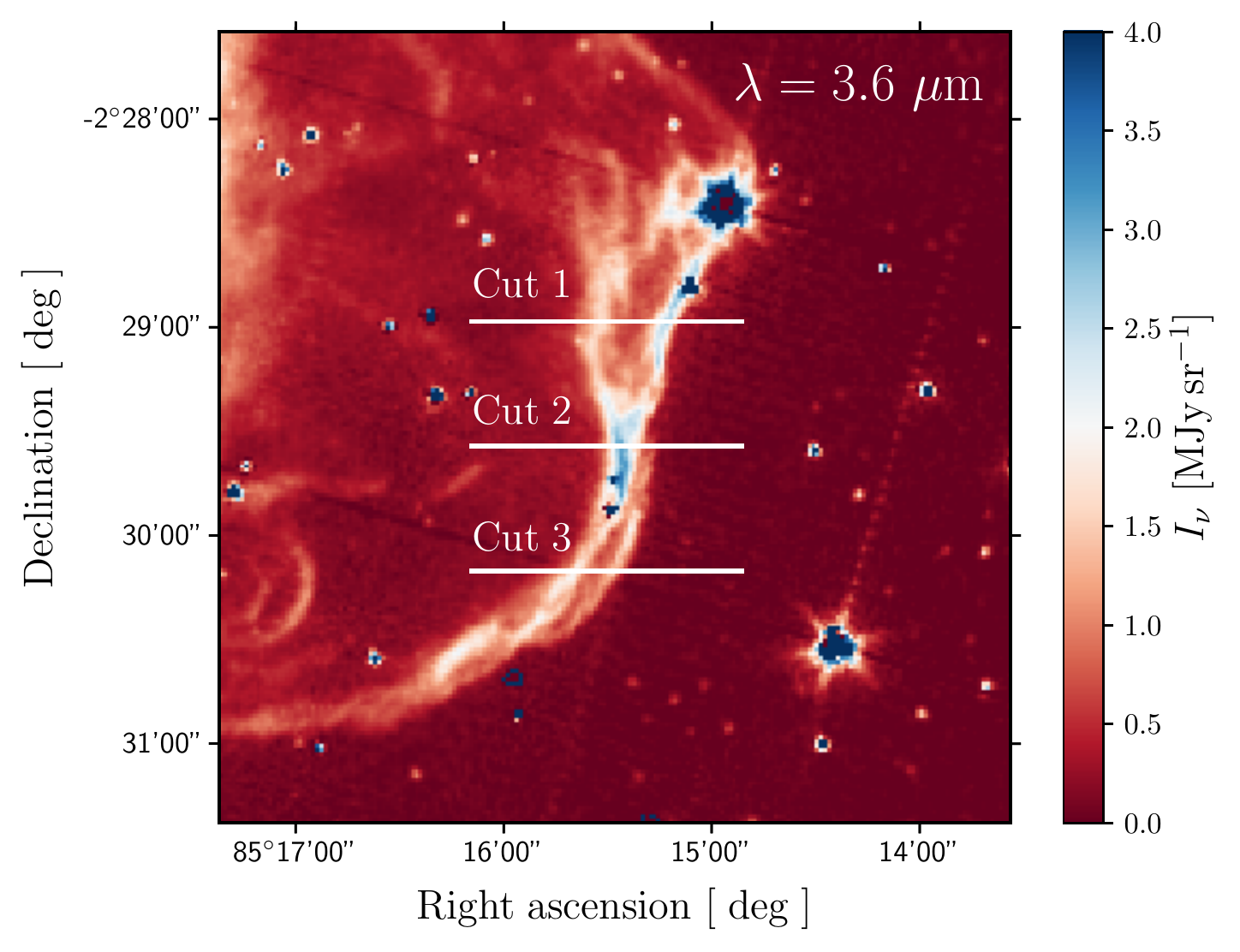}
	\includegraphics[width=0.5\textwidth, trim={0 0cm 0cm 0cm},clip]{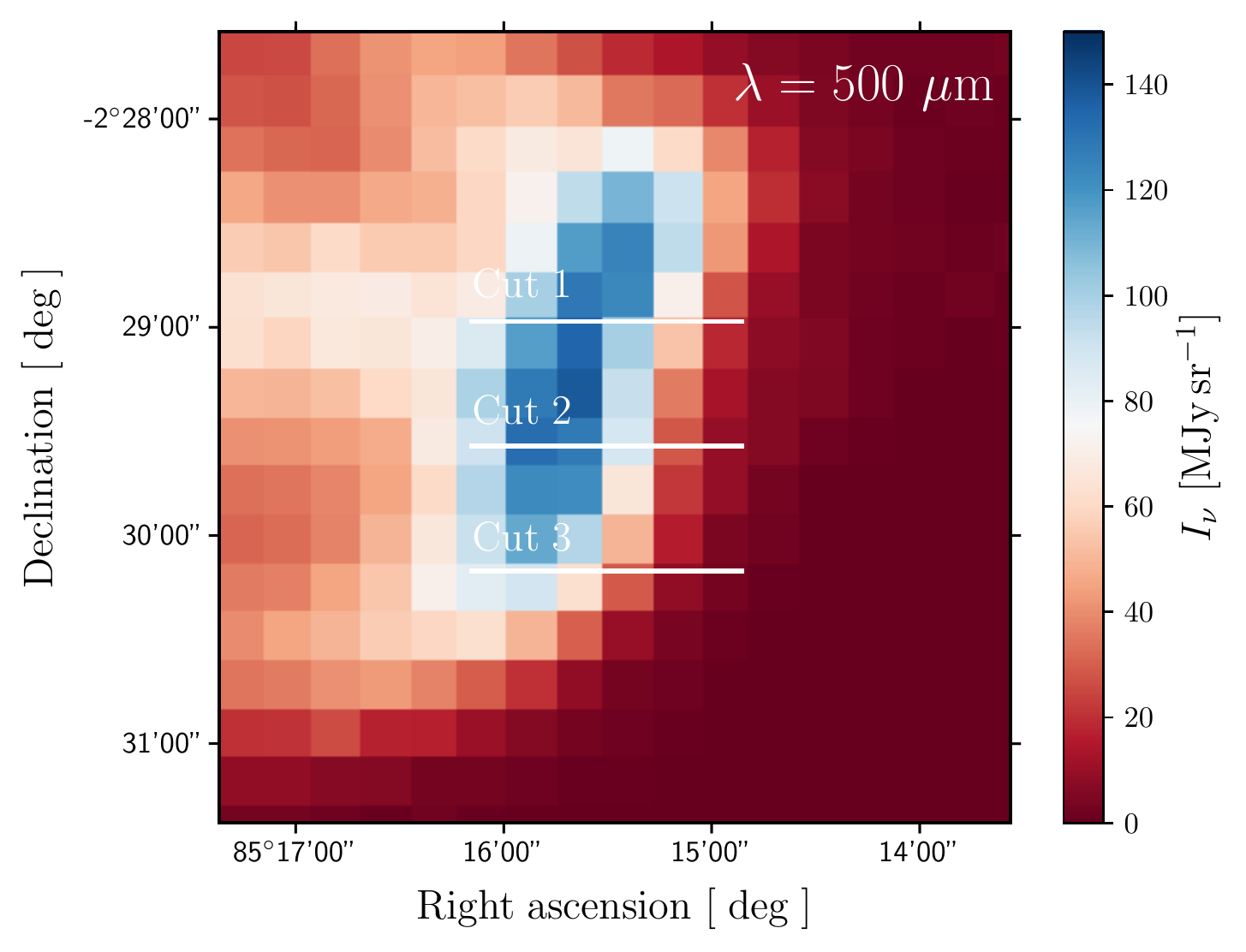}
    \caption{Top: the Horsehead seen with IRAC at 3.6 \mum.
    The three white solid lines correspond to the three cuts we use in our 
    study.
    Bottom: the Horsehead seen with SPIRE at 500 \mum.}
    \label{fig:HH_24}
\end{figure}

\subsection{Observations with Spitzer and Herschel}
\label{sub:sub:observations}
We use \spitzer~and \herschel~observations (see Appendix\,\ref{appendix:HH_obs})
of the Horsehead in
10 photometric bands from 3.6
\mum~to 500 \mum, which cover nearly the entire dust spectrum. The processing
of the \spitzer~maps is detailed in \cite{bowler_infrared_2009}. Data were processed in the HIPE environment, with standard \herschel~corrections for instrumental effects and 
glitches. 
PACS 70 \mum~and 160 \mum~maps were obtained after the superposition 
of two observations with a scan speed of 20\arcsec/s whose directions were
perpendicular to one another. The overall duration of these observations 
is 4122 seconds and cover 8.8\arcmin $\times$~4.5\arcmin~of the Horsehead.
Concerning SPIRE 250 \mum, 350\mum~and 500 \mum, they were obtained after the
superposition of two observations with a scan speed of 30\arcsec/s 
whose directions were perpendicular to one other. The overall duration 
of these observations is 1341 seconds and they cover 8\arcmin $\times$~8\arcmin~of the Horsehead.
Striping induced by offsets in the flux calibration from one detector to
another was removed using the Scan Map Destriper module included in the HIPE
environment.

We study the observed emission profiles through three different cuts across the
Horsehead (see Fig.\,\ref{fig:HH_24}). The calibration uncertainty in the IRAC bands 
(\IRACun, \IRACdeux, \IRACtrois~and \IRACquatre) is 2 $\%$
\citep{reach_absolute_2005}, 4 $\%$ in \MIPSun~\citep{engelbracht_absolute_2007}, 
5 $\%$ in \MIPSdeux~\citep{gordon_absolute_2007}, 12 $\%$ in
\MIPStrois~\citep{stansberry_absolute_2007} and 15 $\%$ in
the 3 SPIRE bands 
\citep{swinyard_-flight_2010}. 
In this study, we considered all these errors to be independent of the wavelength 
to first order. Also, we consider that the emission in all of these 10 
bands is coming from dust, which is not completely the case in \IRACun~
and \IRACdeux. We estimate with a model of atomic and
molecular gas in PDRs, the Meudon PDR Code
\citep{le_petit_model_2006}, that gas can contributes less than 10$\%$ of the
flux. However, 
this contribution does not affect the bulk of our results hence we 
consider that the observed emission is dust emission. Nevertheless,
one must be careful in the interpretation of the observations as gas emission
can be larger than dust emission in photometric bands covering shorter 
wavelengths (HST or NIRCAM onboard the JWST).
\begin{figure}
    \centering
    \includegraphics[width=0.5\textwidth, trim={0 0 0cm 0cm},clip]{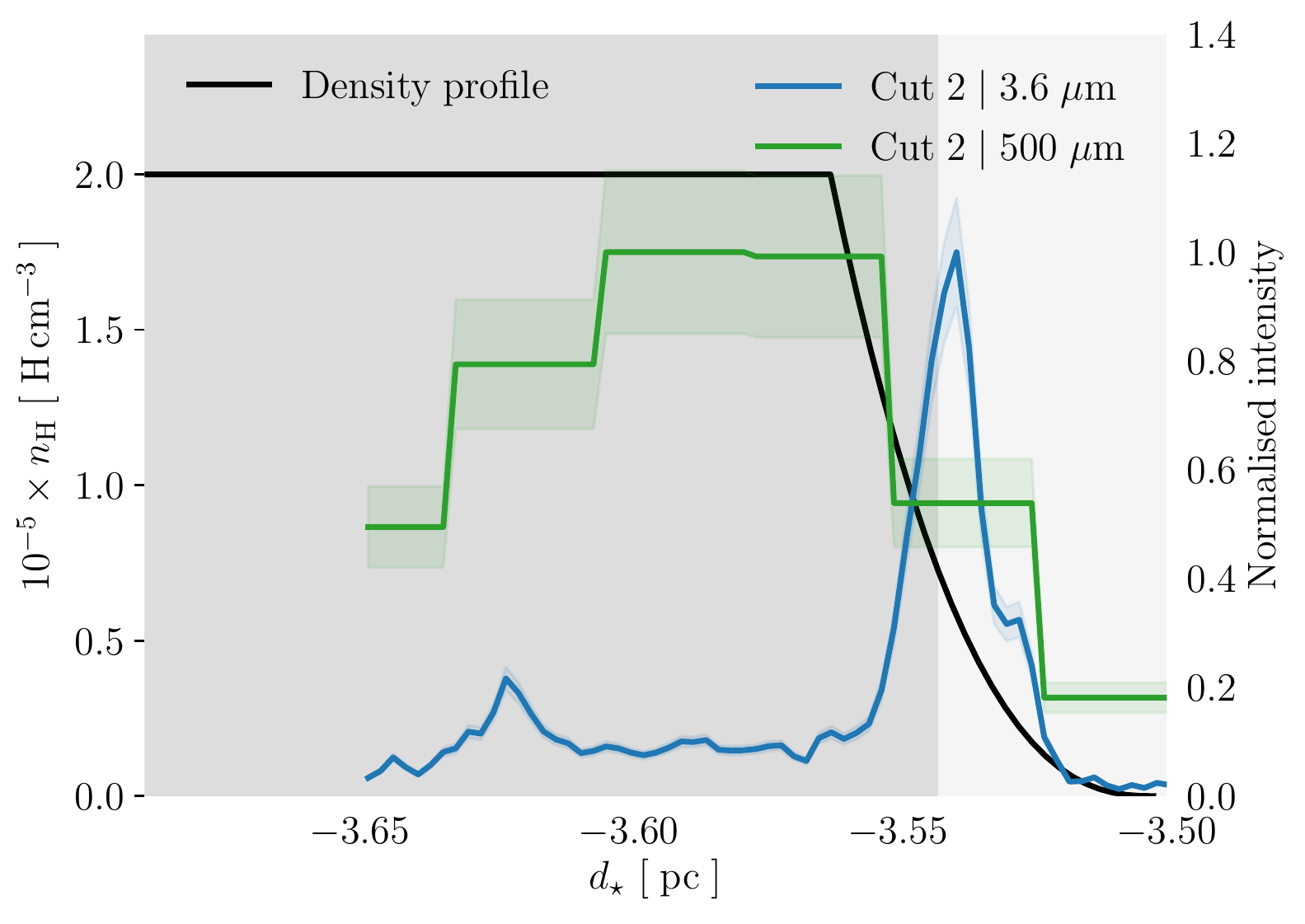}
    \includegraphics[width=0.5\textwidth, trim={0 0 0cm 0cm},clip]{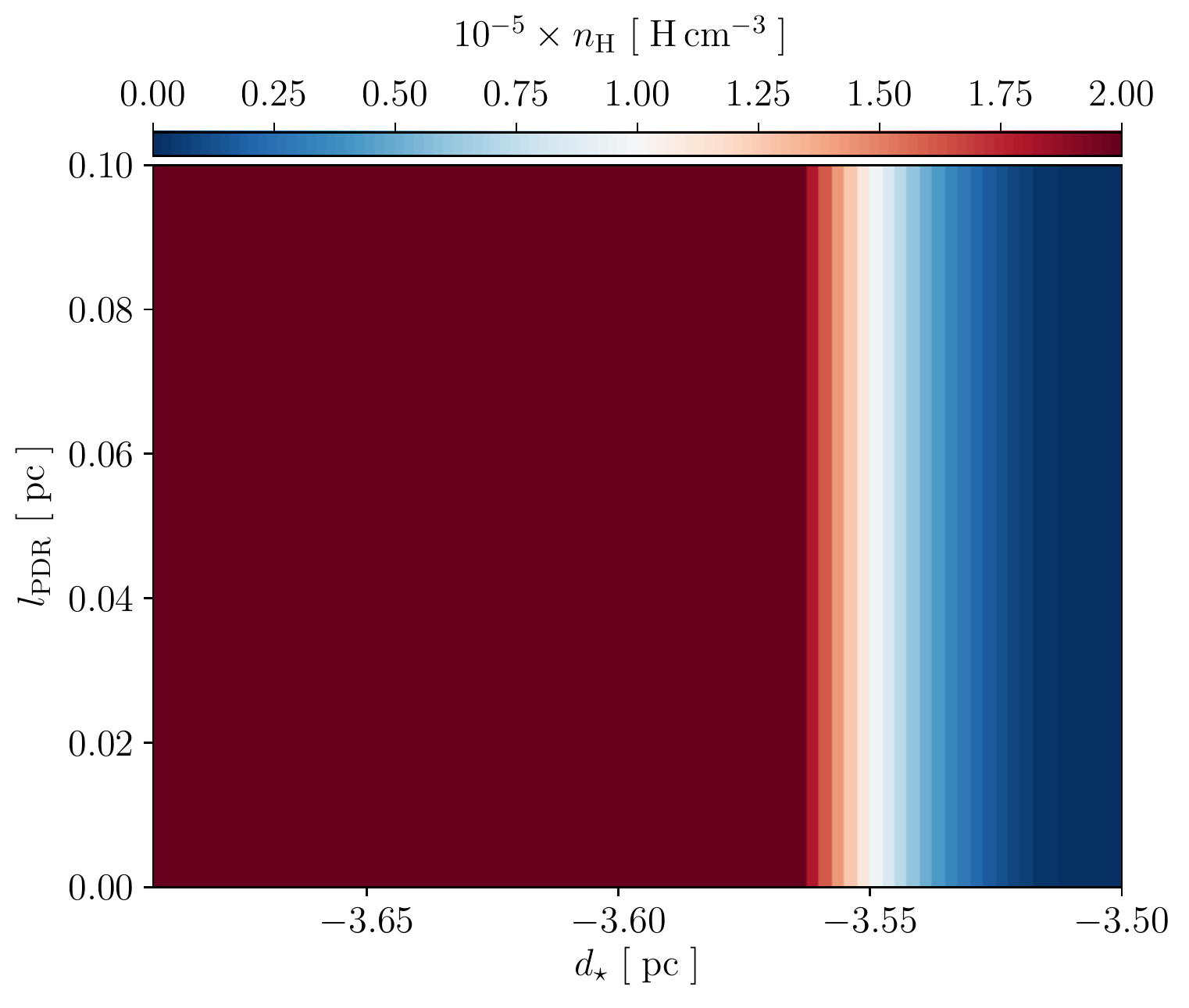}

    \caption{Top: assumed density profile across the Horsehead (black line, see
    Sect.\,\ref{sec:sub:density_profile}).
    Normalised observed dust emission (blue line) in \IRACun~(see Fig.\,\ref{fig:HH_24}). Normalised dust observed emission (green line) in \SPIREtrois. The grey
    part corresponds to the inner part of the Horsehead, defined in
    Sect.\,\ref{sec:sub:evolved_grains}. Bottom: density profile in the 2D-space defined by the distance to
    the star, $d_{\star}$, and the length of the Horsehead along the line of sight, \lpdr.}
    \label{fig:density_profile}
\end{figure}

\subsection{Density profile across the Horsehead}
\label{sec:sub:density_profile}
In this paper, radiative transfer calculations are performed, which 
require information on the density profile across the Horsehead.
We use the profile described in \cite{habart_density_2005}. As the H$_{2}$ 1-0 S(1) fluorescent emission is very sensitive to both the radiation
field and the gas density, they observed this line with the SOFI 
instrument at the NTT. 
This observation was combined with previous observations of H$_{\alpha}$ and dust mid-IR emission in order to
constrain
the density profile at the edge of the Horsehead. \cite{habart_density_2005} also used CO mm observations from
the IRAM 30-m telescope
\citep{abergel_isocam_2003,teyssier_carbon_2004}
and the Plateau de Bure Interferometer \citep{pety_are_2005}
as well as 1.2 mm dust continuum emission obtained with MAMBO at the IRAM 30-m telescope
\citep{teyssier_carbon_2004} to constrain the density profile in the inner part. All these observations were interpreted with the Meudon PDR Code. This density profile (see Fig.\,\ref{fig:density_profile}, upper panel) was also used in \cite{compiegne_dust_2008} and \cite{arab_evolution_2012-2} and is defined as follows : 

\begin{equation*}
    \label{eq:density_profile}
    n_{\mathrm{H}}(z)=\left\{ 
\begin{array}{l l}
  n_{0} \times \left(\frac{z}{z_{0}}\right)^{\gamma}  & \quad \text{if $z<z_{0}$}\\
  n_{0} & \quad \text{if $z>z_{0}$}\\ \end{array} \right.
\end{equation*}
where :
\begin{equation*}
    \label{eq:density_parameters}
    n_{0} = 2 \times 10^{5}\,\mathrm{H\,cm^{-3}} \; ; \; 
    z_{0} = 0.06\,\mathrm{pc} \; ; \; 
    \gamma = 2.5 \; ; \; z = d_{\star} - d_{\mathrm{edge}}.
\end{equation*}
with $z$ the position from the edge of the Horsehead, $\gamma$ the 
power-law exponent of the gas density profile and $z_0$ the depth
beyond which constant density $n_0$ is reached. 

In this study they also estimated the length of the Horsehead along the line
of sight, \lpdr. They
found that this parameter is constrained to be between 0.1 pc and 0.5 pc. We assume
that the density profile is independent of the position along the line 
of sight (see Fig.\,\ref{fig:density_profile}, bottom panel).

\section{Dust modelling}
\label{sec:models_tools}
The interpretation of the multi-wavelength observations of the 
Horsehead depends on its structure, the incident radiation field
and the dust model. We therefore need a dust model and modelling tools 
to compute dust emission based on the local physical conditions. First,
we describe our adopted dust model THEMIS;
second, we introduce DustEM, that is used to compute the local dust
emission and we describe how dust emission evolves with its properties
in the optically thin case using DustEM.

\subsection{THEMIS}
\label{sec:sec:THEMIS}

The Heterogeneous dust Evolution Model for Interstellar Solids, THEMIS\footnote{THEMIS is available here : \href{https://www.ias.u-psud.fr/themis/index.html}{https://www.ias.u-psud.fr/themis/}}
\citep[e.g.,][]{jones_evolution_2013, kohler_hidden_2014, jones_global_2017}, 
is based on observational constraints and laboratory measurements on interstellar
dust analogues that are amorphous hydrocarbons, a-C(:H);
\citep[e.g.,][]{jones_variations_2012-2, jones_variations_2012-1, jones_variations_2012} and amorphous silicates, a-Sil.
This model includes dust evolution through 
processes such as photo-processing, fragmentation and coagulation resulting
from wide variations in the ISM physical condition. 

THEMIS for the diffuse ISM \citep{jones_evolution_2013, kohler_hidden_2014, ysard_dust_2015} is composed of amorphous silicates (\py) surrounded by a mantle of
aromatic-rich carbon, and amorphous hydrocarbon
solids which encompasse a-C:H material that are H-rich hence aliphatic-rich 
and a-C material that are H-poor hence aromatic-rich. Assuming a typical penetration
depth of a UV photon in an amorphous carbon 
grain is about 20 nm \citep[see Fig.\,15 in][]{jones_variations_2012-1},
carbonaceous grains that are smaller than 20 nm are entirely photo-processed
hence aromatic. Larger grains are composed of an aliphatic core and surrounded by 
an aromatic mantle that
is 20 nm thick, which prevents photo-processing
of the core hence allows the core to remain aliphatic. 
This view provides us with a continuous description of carbonaceous grains
from the smallest that mostly contain aromatic cycles and are stochatiscally heated to 
the largest that are at thermal equilibrium. Details about the size distribution
can be found in table \ref{tab:parameters_size_distribution}. As these grains are composed of either an a-C:H core or a silicate core 
surrounded in both cases by an aromatic carbonaceous mantle, they are called Core-Mantle 
grains (CM).

In the dust evolution framework assumed by THEMIS \citep{jones_cycling_2014}, 
large grains can form a second
mantle either through accretion of C and H atoms, available in the gas
phase or through coagulation of a-C nano-grains on the larger grains surfaces. These
grains
are called Core-Mantle-Mantle (CMM). In denser regions, CMM grains coagulate
together to form aggregates \citep{kohler_dust_2015} called Aggregate-Mantle-Mantle (AMM) grains. Where the shielding from energetic photons is efficient enough, a mantle of water
ice can form
around AMM, leading to Aggregated-Mantle-Mantle-Ice (AMMI) grains.

\begin{figure}[h]
\centering
	\includegraphics[width=0.5\textwidth, trim={0 0cm 0cm 0cm},clip]{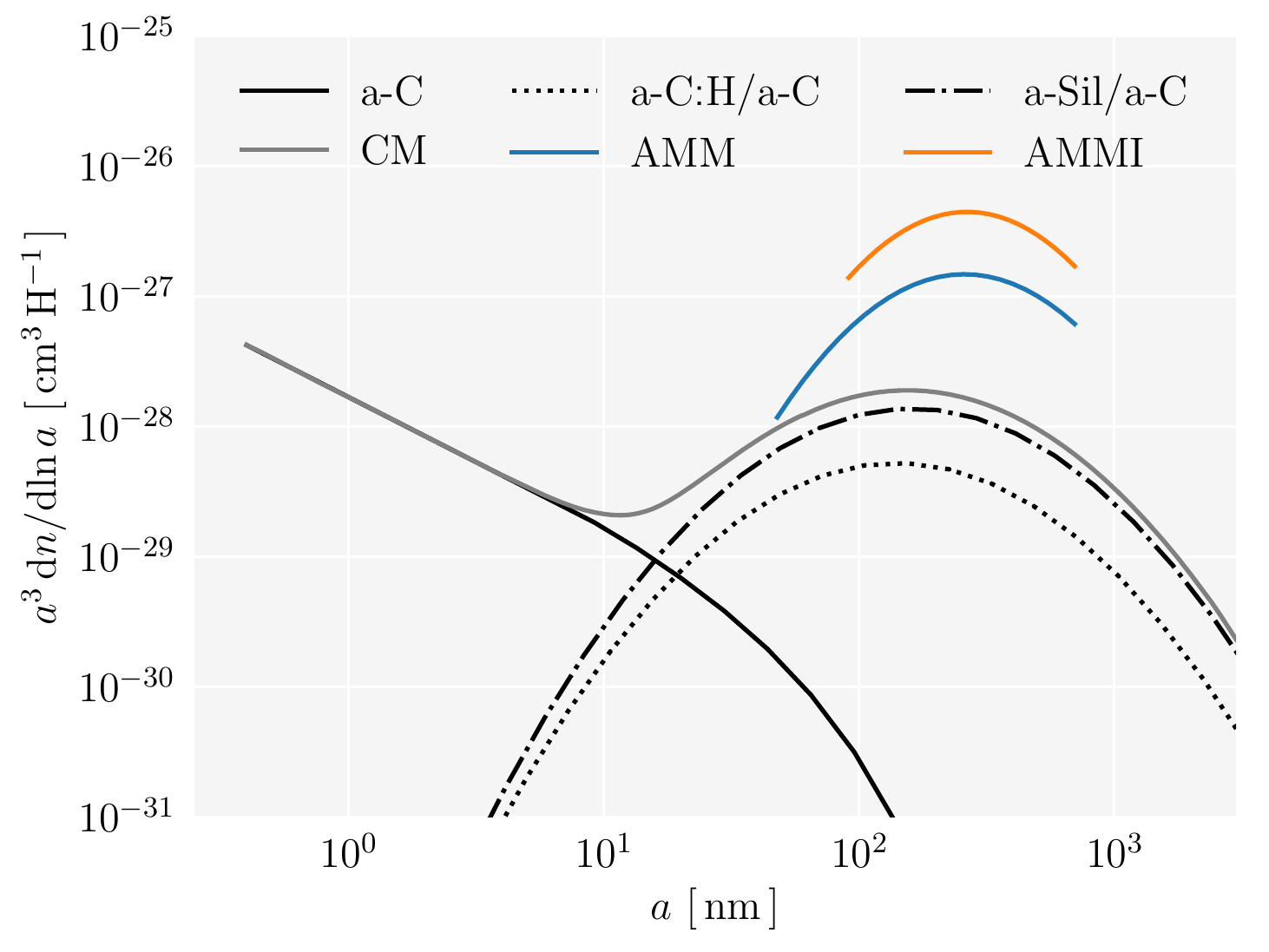}
	\includegraphics[width=0.5\textwidth, trim={0 0 0cm 0cm},clip]{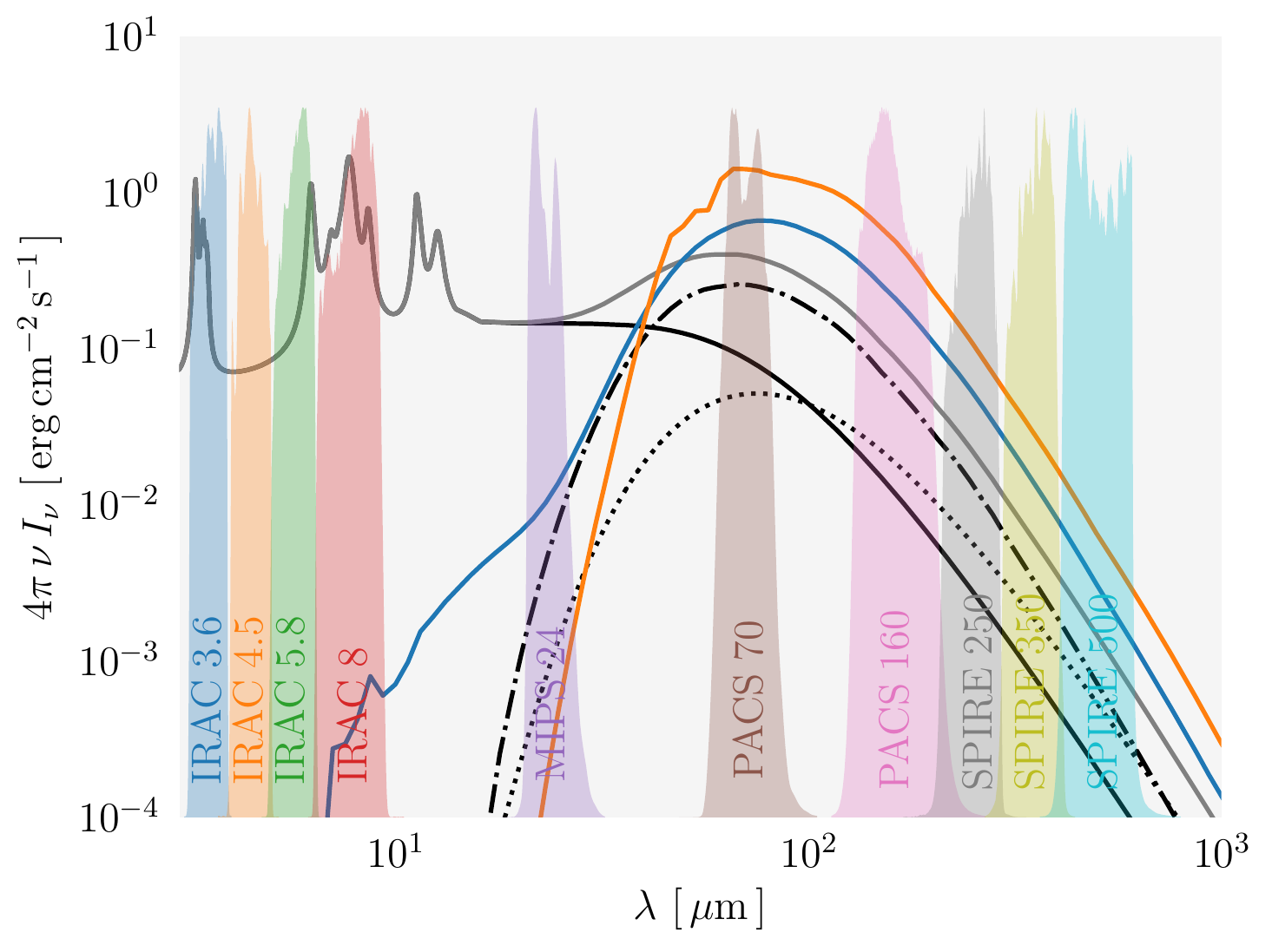}
    \caption{Top : Size distributions of the dust mixtures from THEMIS 
    (parameters are listed in Table.\,\ref{tab:parameters_size_distribution}) for CM-grains (grey line), AMM (blue line) and AMMI (orange line). Black line, dotted-line and dash-dot line 
    correspond to \cm, a-C:H/a-C and a-Sil/a-C respectively. Bottom :
    associated spectra, computed with DustEM with a radiation field corresponding to a star at 34\,600 K with
    \G~= 100.}
    \label{fig:s_dist}
\end{figure}

In the following, we use several dust mixtures \citep[see Fig.\,1 in][]{jones_global_2017}. Parameters associated with the size distribution
of these dust mixtures can be found in Table\,\ref{tab:parameters_size_distribution} and the size distributions themselves in Fig.\,\ref{fig:s_dist} (upper panel) with the 
associated spectra (see Fig.\,\ref{fig:s_dist}, 
bottom panel), computed with DustEM (see Sect.
\ref{sec:sec:DustEM}). In the near-IR (1 to 5 \mum) and mid-IR (5 to 30 \mum), dust
emission comes mainly from the \cm~grains. In the far-IR (from 50 to 500 \mum), dust
emission comes mainly from \py~and \CM~grains.

\subsection{Influence of dust properties on its emission with DustEM}
\label{sec:sec:DustEM}
DustEM\footnote{DustEM is available here : \href{http://www.ias.u-psud.fr/DUSTEM}{http://www.ias.u-psud.fr/DUSTEM}} \citep{compiegne_global_2011} is a modelling tool that computes the extinction, the emission and the
polarisation of interstellar dust grains heated by photons, in the
optically thin case (i.e. no radiative transfer). 

In order to disentangle the effects of radiative transfer from variations in the dust properties on emission, we study the influence of such variations with DustEM. We modify the following parameters : 

\begin{enumerate}
    \item The \cm~abundance, i.e. the \cm~mass to gas ratio, \abvsg, 
    varying from 0.01 $\times\,10^{-2}$ to 0.20 $\times 10^{-2}$ with 
    steps of 0.01 $\times\,10^{-2}$.
    \item The \cm~minimum size, \aminvsg, varying from 0.4 nm to 0.9 nm
    with steps of 0.02 nm.
    \item The slope of the \cm~power law size distribution, $\alpha$, 
    varying from -6 to -4 with steps of 0.1.
\end{enumerate} 

The results are shown in Fig.\,\ref{fig:test} where the spectra in 
panels \textit{d}, \textit{e} and \textit{f} are associated with the size distributions 
in panels \textit{a}, \textit{b} and \textit{c}, respectively. All the spectra are obtained with a radiation field 
that corresponds to a blackbody at 34600 K scaled so that $G_0=100$ (i.e. 
the radiation field illuminating the Horsehead).

A decrease in \abvsg~or an increase in \aminvsg~or 
$\alpha$ leads to a decrease in the smallest \cm~grains hence
a decrease in the near-IR emission. As the total dust mass is fixed, 
an increase in \aminvsg~or $\alpha$ leads to a redistribution of
the dust mass from the smallest to the largest \cm~grains hence an 
increase in the mid-IR emission. In the far-IR, dust emission is
unaffected by variations in \abvsg, \aminvsg~and $\alpha$ as 
\cm~grains are barely responsible for any dust emission at these 
long wavelengths. However, the dust emission in the
far-IR slightly increases with an increase in $\alpha$ as the mass of the largest \cm~increases significantly, unlike an increase in \aminvsg. 

\begin{figure}[t]
\centering
	\includegraphics[width=0.5\textwidth, trim={0.0cm 0cm 0.0cm 0.0cm},clip]{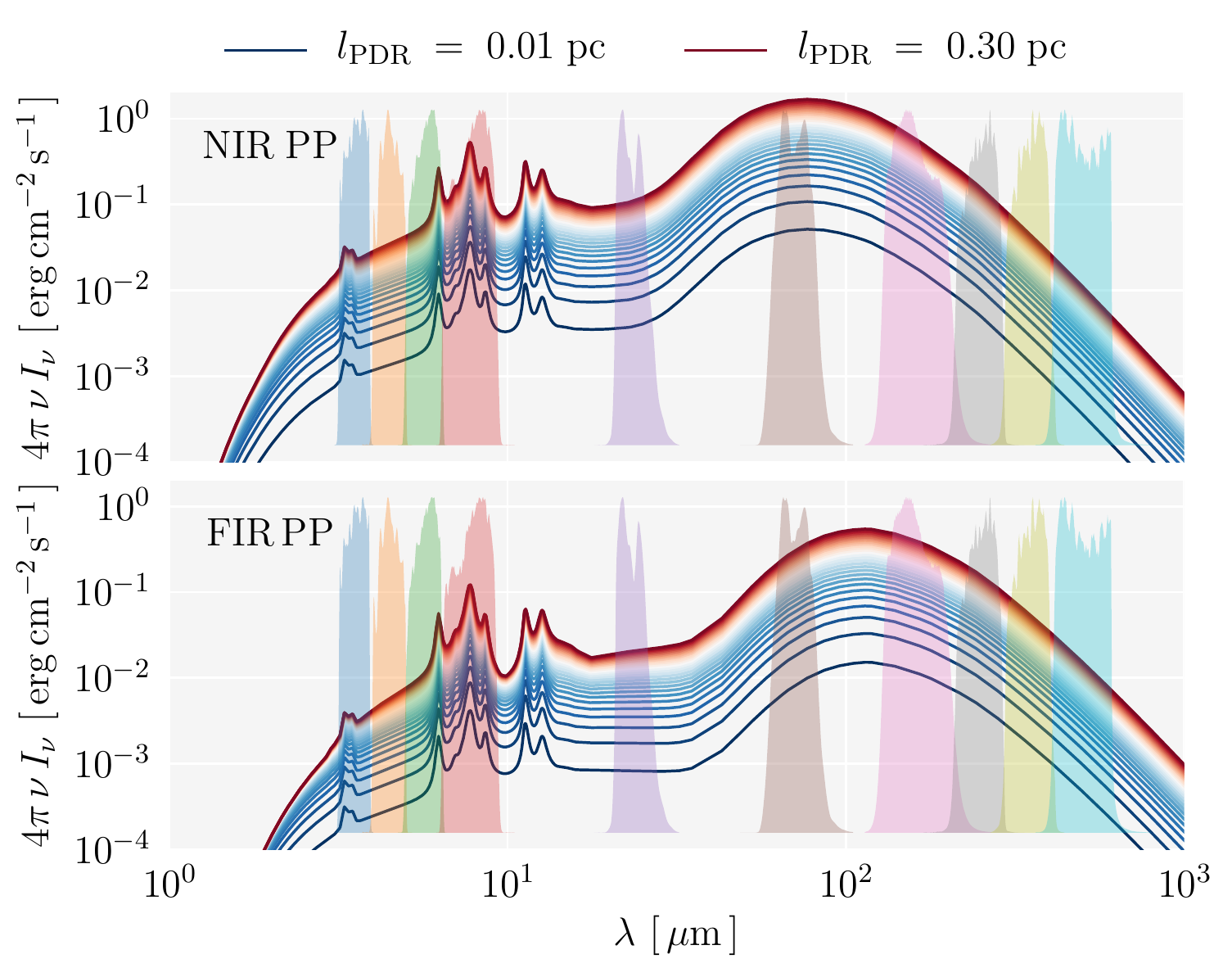}
	\includegraphics[width=0.5\textwidth, trim={0.0cm 0cm 0.0cm 0.0cm},clip]{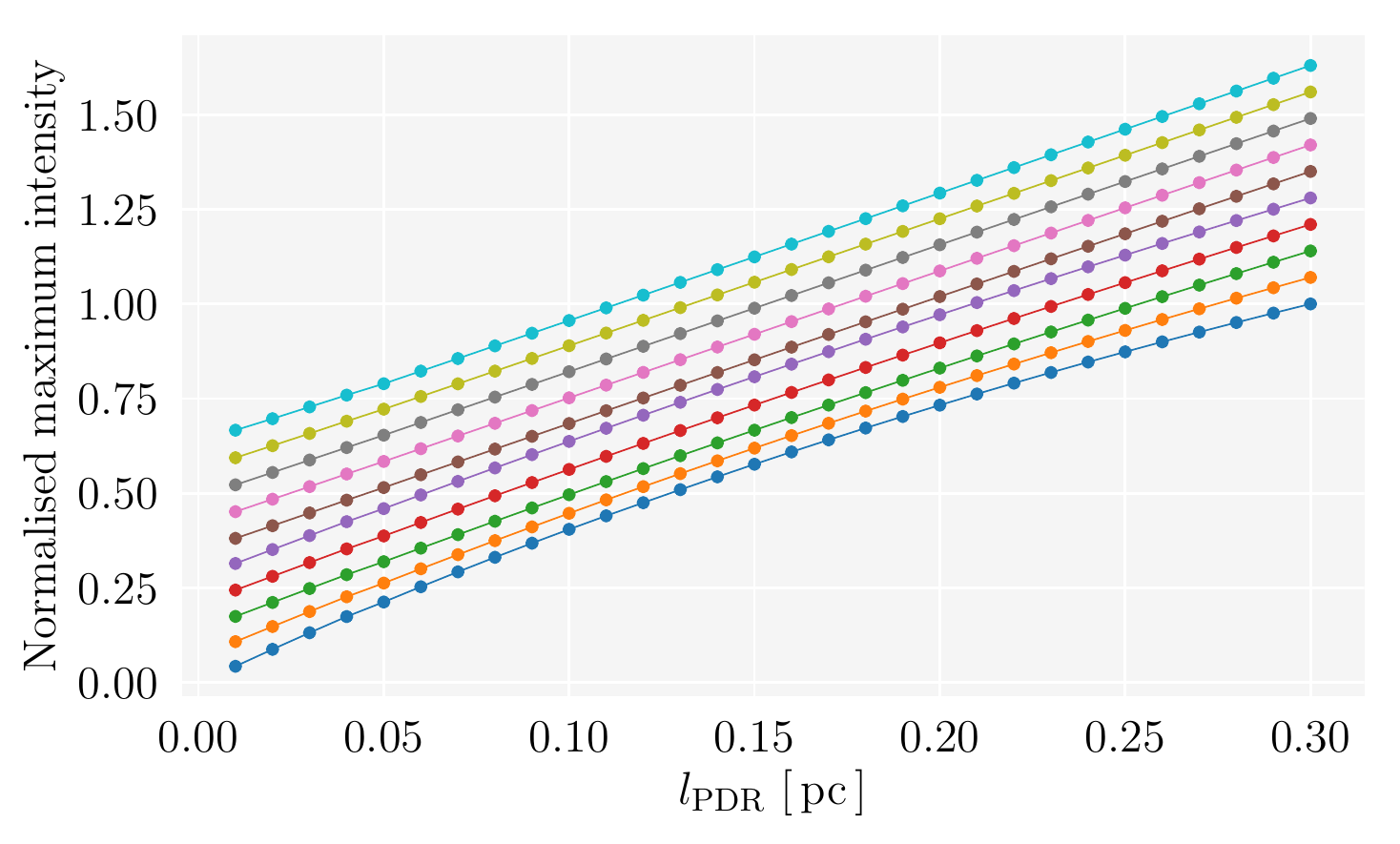}
    \caption{Top: Dust modelled spectra with SOC at the near-IR peak position (NIR PP) for \lpdr~varying from 0.01 pc (blue line) to 0.30 pc 
    (red line) with a step of 0.01 pc. The 10 photometric bands are 
    shown in colours behind the lines. Middle: same at the far-IR peak position (FIR PP). Bottom: Normalised
    $I_{{\mathrm{mod,\,max}}}(i)$ for each band as a function of 
    \lpdr. Colours refer to the different photometric bands shown in
    the upper panels. Lines are shifted for clarification (from top to bottom
    in the order of decreasing wavelength).}
    \label{fig:I_emi_profile_1}
\end{figure}

\section{Radiative transfer modelling within the Horsehead}
\label{sec:dust_emission_radiative_transfer}
The Horsehead is an optically thick region that requires a radiative transfer
modelling to
properly interpret our multi-wavelength observations. We present the 3D radiative
transfer code SOC we use in this study. Performing radiative transfer
is time consuming, and so 
we here explore the influence of the Horsehead length along the line 
of sight \lpdr, and dust properties (i.e. \abvsg, \aminvsg, $\alpha$) on dust
emission after radiative transfer calculations. 

\subsection{Radiative transfer code : SOC}
\label{sec:sub:sub:SOC}
SOC is a 3D Monte Carlo radiative transfer code, parallelised using
OpenCL libraries \citep{juvela_soc_2019}, that computes dust emission
and scattering. SOC has been benchmarked with other radiative transfer
codes in \cite{gordon_trust._2017} and used in \cite{juvela_dust_2018, juvela_galactic_2018, juvela_synthetic_2019}.

The radiation field corresponds to that of a blackbody at 34\,600 K produced
by a star to which a dilution factor has been applied to obtain \G~= 100 at the Horsehead edge. 
This radiation field is estimated on a logarithmic grid of 334 frequencies 
that extends from $3\times 10^{9}$~Hz to $3\times 10^{16}$~Hz. As the Horsehead 
edge is located outside the HII region, there are no photons above 13.6 eV 
hence we applied the Lyman cut to the radiation field that is heating the 
Horsehead edge. Each frequency is simulated with $10^{6}$ photons.

In SOC, clouds can be defined on regular Cartesian grids or octree grids. In
this paper, we model the Horsehead using a cartesian grid that contains 
$N_{\mathrm{X}} \times N_{\mathrm{Y}} \times N_{\mathrm{Z}}$ cubes that measure 
0.0025 pc per side. 
$N_{\mathrm{X}}$ is equal to 77 and corresponds to the number of cubes 
along the Horsehead-star axis. $N_{\mathrm{Y}}$ is equal to 7 and corresponds 
to the number of cubes along the axis perpendicular to the Horsehead-star axis 
and the line-of-sight axis (i.e. the observer-Horsehead axis). $N_{\mathrm{Z}}$
correponds to the number of cubes in the Horsehead along the line-of-sight hence
depends on the value of \lpdr: $N_{\mathrm{Z}}=$ \lpdr\,/\,0.0025 pc. For each 
cube, we associate a value of gas density as described in
Sect.\,\ref{sec:sub:density_profile}.

In our study, 
we compute only dust emission as 
regardless of the dust properties, dust scattering contributes up to less than 
1 $\%$ to the total dust brightness in the near-IR photometric bands. After the integration along the line-of-sight, dust emission profiles across
the Horsehead are integrated into the different photometric bands and convolved
with the PSFs. 

\subsection{Influence of \lpdr~on dust emission}
\label{sec:sub:sub:lpdr_radiative_transfer}
In the following, we study dust emission at two positions : the near-IR
peak position (NIR PP) in the Horsehead and the far-IR peak position (FIR PP). These positions
corresponds respectively to the peak of emission in \IRACun~and \SPIREtrois,
shown in the Fig.\,\ref{fig:density_profile}. Also, in order to lighten the reading of the results obtained, we introduce $I_{{\mathrm{mod,\,max}}}(i) = \mathrm{max}\left(I_{\mathrm{mod},\,i}(z)\right)$
where $I_{\mathrm{mod},\,i}(z)$ is 
the dust modelled emission in the $i$-th band at the position $z$ along the cut.

Whether it is at the NIR PP or at the FIR PP,
dust emission increases in all bands with \lpdr~(see Fig.\,\ref{fig:I_emi_profile_1},
top and middle panels) since the dust mass increases along the line
of sight as the column
density\footnote{$N_{\mathrm{H}}(z)=n_{\mathrm{H}}(z)\,l_{\mathrm{PDR}}$} 
increases with \lpdr. One may also note that dust emission increases
linearly with \lpdr~(see Fig.\,\ref{fig:I_emi_profile_1},
bottom panel) revealing that dust self-absorption, which depends on both the column
density and the wavelength, is negligible at these
wavelengths in the \lpdr~range we are considering. Consequently, we can consider that
the
intensity increases linearly with \lpdr~in the near, mid and far-IR and does
not affect the shape of the dust spectrum. In the following, we therefore consider \lpdr~as a multiplying factor
on the dust spectrum.

\subsection{Influence of dust properties on dust emission after radiative
transfer}
\label{sec:sub:dust_prop_rad_transfer}
\begin{figure*}[!h]
    \includegraphics[width=1\textwidth, trim={0 0 0 0},clip]{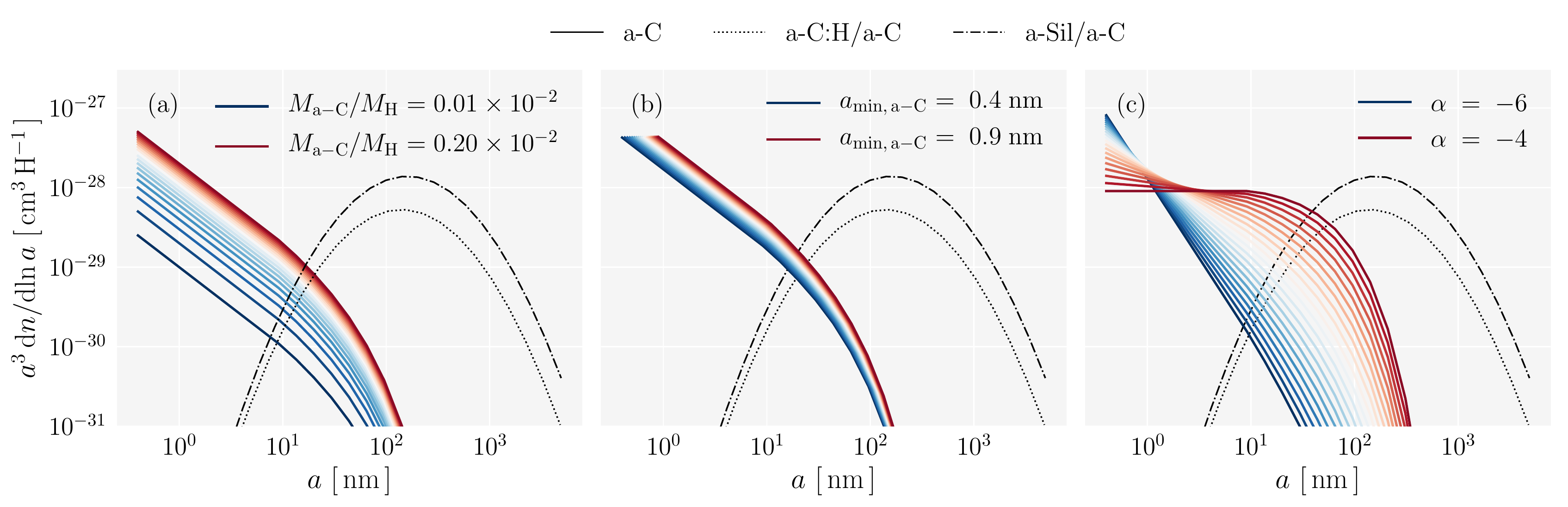}
    \includegraphics[width=1\textwidth, trim={0 0 0 0},clip]{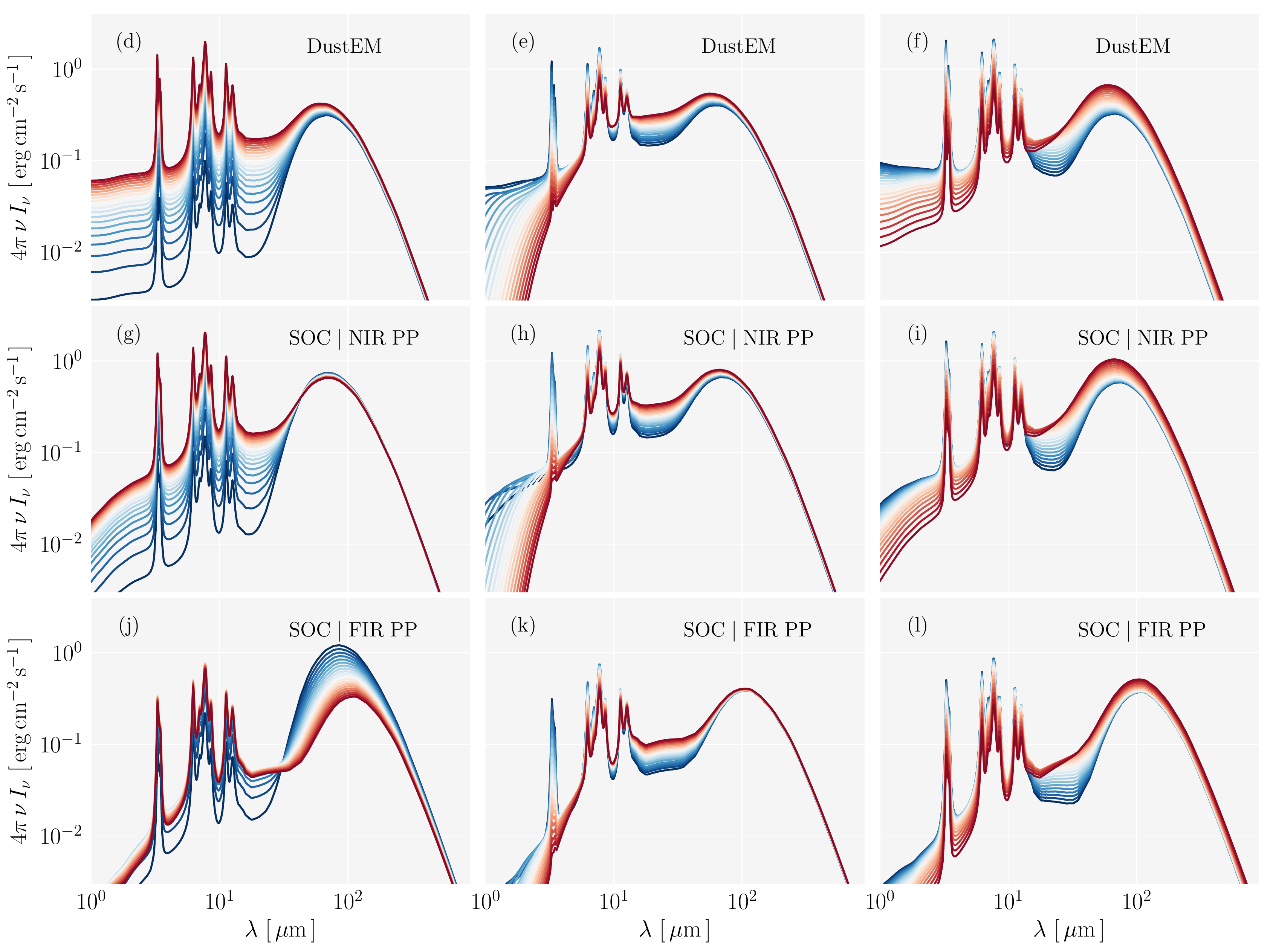}
    \caption{
    Dust size distributions for different values of \abvsg~(panel a), \aminvsg~(panel b) and $\alpha$ (panel c). Panels d-l show dust spectra for the size distributions shown in the corresponding top-row panels. The spectra are computed with DustEM (panels d-f) and with radiative transfer for the NIR (panels g-i) and the FIR (panels j-l) peak positions.}
    \label{fig:test}
\end{figure*}

Conversely to Sect.\,\ref{sec:sec:DustEM} where we study
the influence of dust properties on dust emission in the optically thin
case, we study here the influence of these properties in the optically
thick case by performing a radiative transfer calculation. These results
are shown in Fig.\,\ref{fig:test} where the spectra in panels \textit{g}/\textit{j} , \textit{h}/\textit{k} and
\textit{i}/\textit{l} are respectively associated with the size distributions in panel \textit{a}, \textit{b}
and \textit{c}.
Spectra in panels \textit{g}, \textit{h} and \textit{i} are located at the NIR PP and
those in panels \textit{j}, \textit{k} and \textit{l} are located at the FIR PP.

Dust grains are warmer at the NIR PP
 (see Fig.\,\ref{fig:test}, panels \textit{g}, \textit{h} and \textit{i})
than at the FIR PP (panels
\textit{j}, \textit{k} and \textit{l}) as the
maximum intensity shifts towards longer wavelengths.
 This effect is due to the damping of the radiation field with increasing depth into 
 the Horsehead. 

One may note that at the NIR PP, dust emission in the far-IR does not vary
with \abvsg~(see
Fig.\,\ref{fig:test}, panel \textit{g}), conversely to that is seen in the inner part
(panel \textit{j}). As dust emission 
in the far-IR is unaffected by variations in \abvsg~in the optically thin case 
(see Sect.\,\ref{sec:sec:DustEM}), this is strictly a radiative transfer effect. As the 
\cm~grains bear a large fraction of the total dust cross-section, an increase in
\abvsg~increases significantly the extinction. Therefore as \abvsg~increases, the radiation field is increasingly damped at
the NIR PP and
there are less photons available at the FIR PP to heat the larger grains. Indeed, as we can
see in panel \textit{j}, the wavelength associates with the maximum of emission shifts towards longer wavelengths with an increase in \abvsg. Therefore, 
dust emission in the far-IR varies with \abvsg, 
due to radiative transfer effects. 

Regarding the other changes in the spectra,
they are due to variations in dust properties and are explained in
Sect.\,\ref{sec:sec:DustEM}. 

\subsection{With evolved grains}
\label{sec:sub:evolved_grains}
Previously, we used only CM-grains throughout the 
Horsehead. To study the influence of dust evolution on the emission across the
Horsehead, we use CM-grains with modified size distributions (i.e. CM-grains with values of
    \abvsg, \aminvsg~and $\alpha$ that differ from the diffuse ISM) in the outer
    part of the Horsehead where the dust is likely
to be more diffuse ISM-like, and aggregate-grains (AMM, AMMI) above a density threshold of 7 $\times\,10^{4}$ H.cm$^{-3}$, where dust grains
are assumed to be coagulated. In order to simplify our study, we define 3 different cases depending
on the dust we use :
\begin{description}
    \item[$\bullet$ \textbf{Case \textit{a} :}] CM-grains with modified size distributions across all the Horsehead.
    \item[$\bullet$ \textbf{Case \textit{b} :}] CM-grains with modified size distributions in the outer part of the Horsehead and
    AMM in the inner part of the Horsehead.
    \item[$\bullet$ \textbf{Case \textit{c} :}] CM-grains with modified size distributions in the outer part of the Horsehead and
    AMMI in the inner part of the Horsehead.
\end{description}

\begin{figure*}[h]
\centering
	\includegraphics[width=1.\textwidth, trim={0.0cm 0cm 0.0cm 0.0cm},clip]{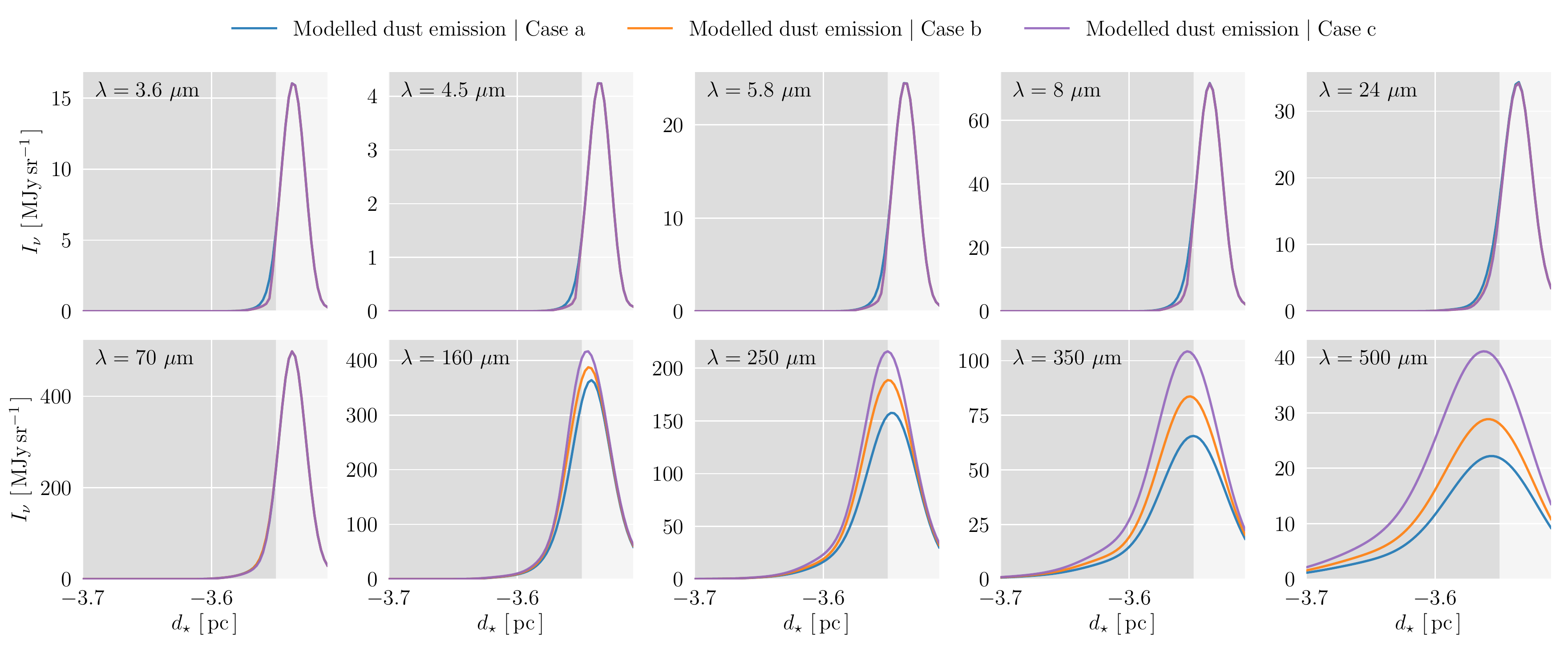}
    \caption{Dust modelled emission profiles in the 10 photometric bands
    for case \textit{a} (blue lines), case \textit{b} (orange lines) and case \textit{c}
     (purple lines). The darker grey parts correspond to the inner
     Horsehead where AMM and AMMI are used in case \textit{b} and \textit{c},
      respectively.}
    \label{fig:I_emi_CM_AMM_AMMI}
\end{figure*}

Dust modelled emission profiles for the three cases are shown in Fig.\,\ref{fig:I_emi_CM_AMM_AMMI}. 

As the maximum of emission in the near and mid-infrared is 
located in the outer part of the Horsehead, there is no modification of dust emission
at these wavelengths as we always use modified CM grains. Regarding dust emission in the
far-infrared, this increases when coagulated (AMM, AMMI) dust grains are used because 
they are more emissive. One may note that AMMI are more emissive than AMM as the dust mass
in AMMI is larger than in AMM because of the ice mantle. 

\section{Comparison with observations}
\label{sec:comparison_observations}
In this section, we constrain our dust model with the observations. First, we present our results using diffuse
ISM-like dust (i.e. CM-grains); second, we introduce the methodology we use in the following parts; third, we constrain the four parameters \abvsg, \aminvsg, $\alpha$ and \lpdr~for the 3 cases of evolved grains as defined in Sect.\,\ref{sec:sub:evolved_grains} 
and across the 3 cuts (see Fig.\,\ref{fig:HH_24}).

\subsection{Diffuse case}
\label{sec:sec:diffuse_case}
\begin{figure*}[h]
\centering
	\includegraphics[width=1.0\textwidth, trim={0.0cm 0.1cm 0.3cm 0.cm},clip]{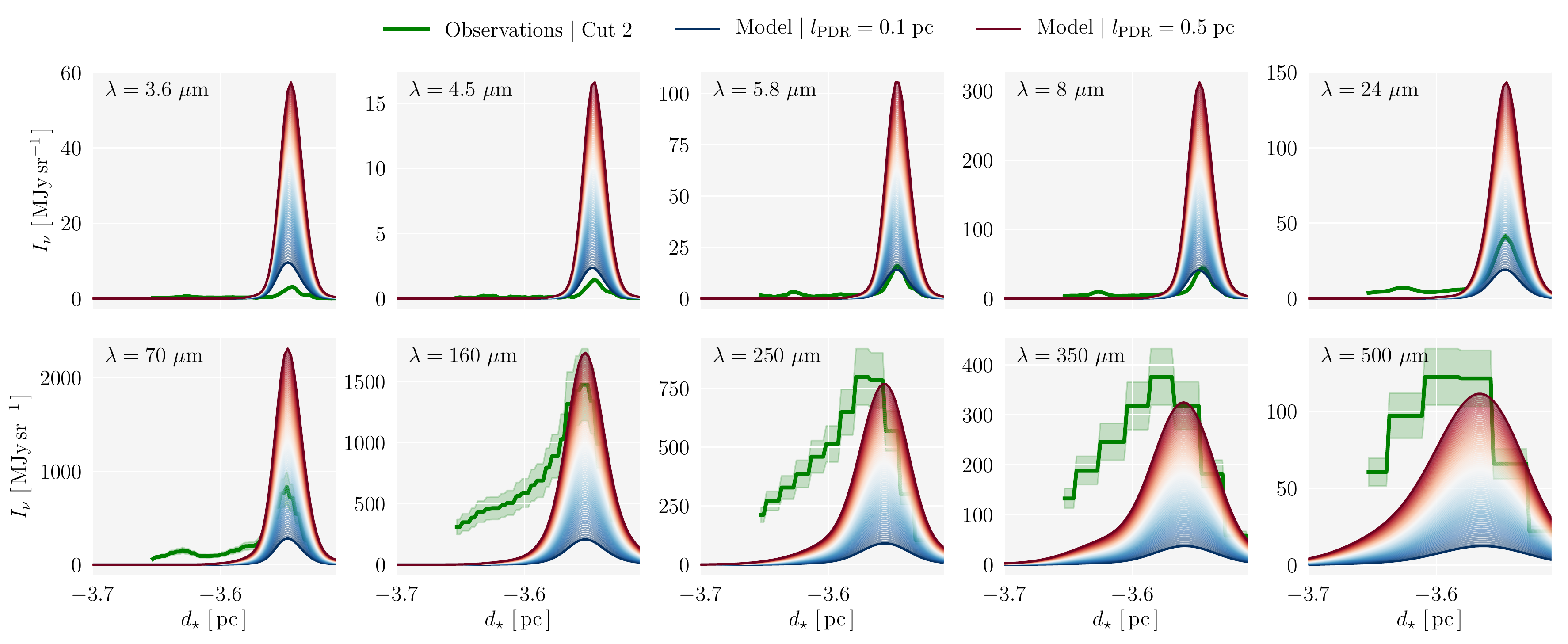}
	\includegraphics[width=1.0\textwidth, trim={0.0cm 0.1cm 0.3cm 0.cm},clip]{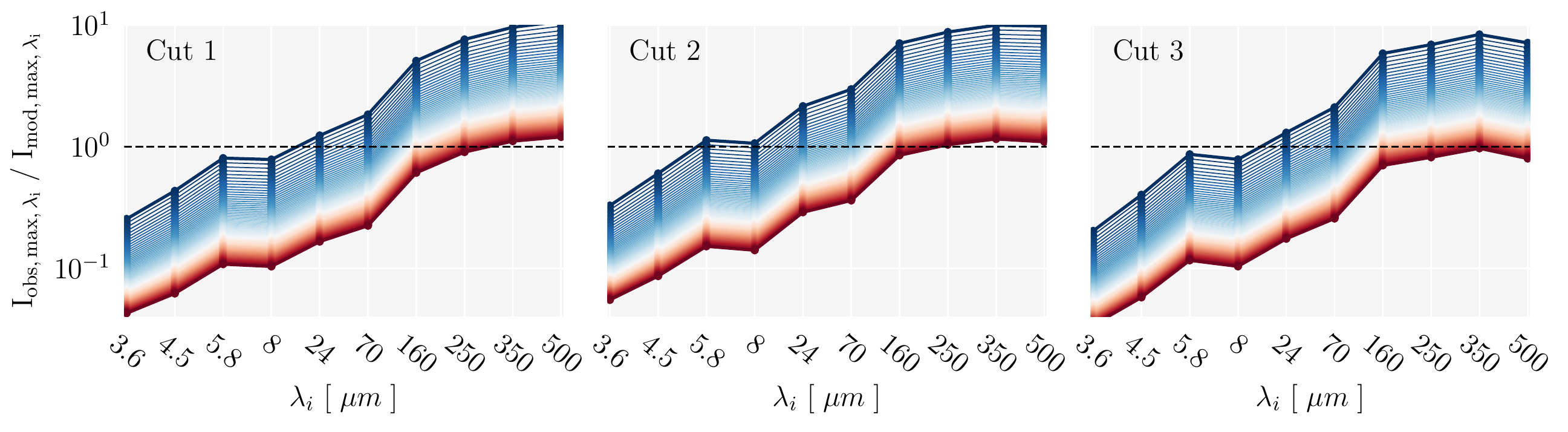}

    \caption{Top : Modelled dust emission for \lpdr~varying from
    0.1 pc (blue lines) to 0.5 pc (red lines) in steps of 0.01
    pc for the 10 photometric bands with diffuse ISM-like dust.
    The observed dust emission for cut 2 is
    shown by green lines. Bottom : ratio between the observed and modelled dust maximum
    emissions in the 10 photometric bands for the three cuts and \lpdr~varying from 0.01 pc (blue lines) to 
    0.5 pc (red lines) with a step of 0.01 pc.
}
    \label{fig:SOC_diffuse}
\end{figure*}

The results are shown in Fig.\,\ref{fig:SOC_diffuse}. The 10 upper panels 
correspond to the modelled emission
across the Horsehead using CM-grains, with \lpdr~varying
from 0.1 pc to 0.5 pc, for the 10 photometric bands. The observed emission is shown for cut 2.
The bottom panels show the corresponding ratios of maximum observed and modelled intensities. 

Regardless of the cut considered, it is not possible to simultaneously
fit the observations in all the photometric bands (see Fig.\,\ref{fig:SOC_diffuse}, upper
panel), whatever the \lpdr~value. With \lpdr~= 0.1 pc,
we are able to roughly reproduce the observations in the near and mid-infrared but in the
far-infrared, the modelled dust emission is too low by a factor $\sim$ 10 (see
Fig.\,\ref{fig:SOC_diffuse}, bottom panels). With \lpdr~= 0.5 pc, we are able to reproduce the observations in the far-infrared but in the near and mid-infrared, the modelled
dust emission is too high by a factor of at least $\sim$ 10.

If \lpdr~is higher than 0.10 pc (see
Sect.\,\ref{sec:sub:density_profile}),
near and mid-infrared modelled
dust emission will always be too high, which implies reducing the dust
abundance that
is responsible for the emission at these wavelengths hence decreasing the \cm~dust-to-gas
ratio, \abvsg~(see Sect.\,\ref{sec:sub:dust_prop_rad_transfer}). On the other hand, the
ratio between the modelled dust emission and the observations
is not the same in the five near and mid-IR bands. One is therefore
required to change the shape of the spectrum by varying \aminvsg~and $\alpha$ (see
Sect.\,\ref{sec:sub:dust_prop_rad_transfer}).

To summarise, it is not possible to reproduce the observations across any of the 
three cuts in the Horsehead for any value of \lpdr, using only diffuse ISM-like dust. We must
therefore consider evolved dust.

\subsection{Methodology}
\label{sub:sub:methodology}
For the sake of reducing computation time, instead of exploring the 4D-space
defined by \abvsg, \aminvsg, $\alpha$~and \lpdr, 
we explore the 3D-space defined by \abvsg, \aminvsg~and
$\alpha$ as variation in \lpdr~does not affect the shape of the dust spectrum (see
Sect.\,\ref{sec:sub:sub:lpdr_radiative_transfer}), conversely to variations in
\abvsg, \aminvsg~and $\alpha$ (see Sect.\,\ref{sec:sub:dust_prop_rad_transfer}). 
Therefore, \lpdr~can be adjusted after the fact.

Adjusting the shape of the modelled dust spectra to the observed dust spectra means
that the ratio between $I_{{\mathrm{obs,\,max}}}(i)$ and $I_{{\mathrm{mod,\,max}}}(i)$
has to be roughly the same
in every band. Therefore, we
minimise the following parameter :
\begin{equation}
    \label{eq:chi2}
    \chi^{2} = \sum_{i\,\in\,\mathrm{filters}} \left(\frac{X_{i}-\mu}{\sigma_{i}} \right)^{2} \;,  
\end{equation}
with
\begin{equation}
    \label{eq:chi2_1}
    X_{i} = \frac{I_{\mathrm{obs,max}}(i)}{I_{\mathrm{mod,max}}(i)} \quad ; \quad 
    \sigma_{i} = r_{\mathrm{obs}}(i)\,X_{i} \quad ; \quad 
    \mu = \left< X_{i} \right>_{i\,\in\,\mathrm{filters}}
\end{equation}
where $r_{\mathrm{obs}}$ is the relative error for each filter and defined in
Sect.\,\ref{sub:sub:observations} and $I_{{\mathrm{obs,\,max}}}(i) = \mathrm{max}\left(I_{\mathrm{obs},\,i}(z)\right)$
with $I_{\mathrm{obs},\,i}(z)$, 
the dust observed in the $i$-th band at the position $z$ 
along the cut.

\vspace{0.3cm}

\noindent The following procedure is thus applied :

\begin{enumerate}
    \item We constrain \abvsg, \aminvsg~and $\alpha$ with a fixed \lpdr~in order
    to adjust the shape of the modelled
    dust spectrum to the observed dust spectrum by minimising \chii.
    \item We use the dust properties (\abvsg, \aminvsg~and $\alpha$) 
    associated with \chimin~(i.e. the minimum value of \chii~in the 3D-space defined by
    \abvsg, \aminvsg~and $\alpha$) and we adjust the 
    overall modelled dust spectrum to the observed dust spectrum by multiplying
    the flux in all bands by the same factor to get $\mu$ = 1, which constrains 
    \lpdr.
\end{enumerate}

We choose to remove the \IRACdeux~and \MIPSdeux~bands because it is not possible
to simultaneously fit
the observations in the 10 bands with these 2 included. We discuss
this decision further in Sect.\,\ref{sec:discrepancy}. 

\subsection{Constrain \lpdr~and dust properties \abvsg, \aminvsg~and $\alpha$}
\label{sec:sec:a_min_vsg_alpha_Eg}
First, we study the \chii~distribution in the 3D-space (\abvsg, \aminvsg, $\alpha$) for each of the 3 cuts and the 3 cases, in order to obtain
the best set of parameters in these 9 cases. The 3D-space is defined as follows :
\begin{enumerate}
    \item \abvsg~varying from 0.001 $\times$ $10^{-2}$ to 0.041 $\times$ $10^{-2}$ with steps of 0.002 $\times$ $10^{-2}$.
    \item \aminvsg~varying from 5 nm to 10 nm with steps of 0.25 nm.
    \item $\alpha$ varying from -13 to -3 with steps of 0.5.
\end{enumerate}
Second, we study the \chii~distribution in the 2D-spaces (\aminvsg, $\alpha$),  
(\abvsg, \aminvsg) and (\abvsg, $\alpha$). Finally, we conclude with the comparison between
the observed and modelled dust emission profiles for each of the 3 cuts and the 3 cases
with the best sets of parameters. 

For more clarity, we define \chimindab~that is the
minimum value of \chii~in the 2D-space (\aminvsg, $\alpha$) for a
given value of \abvsg. We also define \chimin, that is the minimum value of \chii~in the
3D-space (\abvsg, \aminvsg, $\alpha$), i.e. the minimum value of
\chimindab.

\subsubsection{\chii~distribution in the 3D-space (\abvsg, \aminvsg, $\alpha$)}
\label{sec:sec:a_min_vsg_alpha_ab_vsg}
\begin{figure*}[h]
\centering
\includegraphics[width=1.0\textwidth, trim={0.0cm 0.0cm 0.0cm 0.0cm},clip]{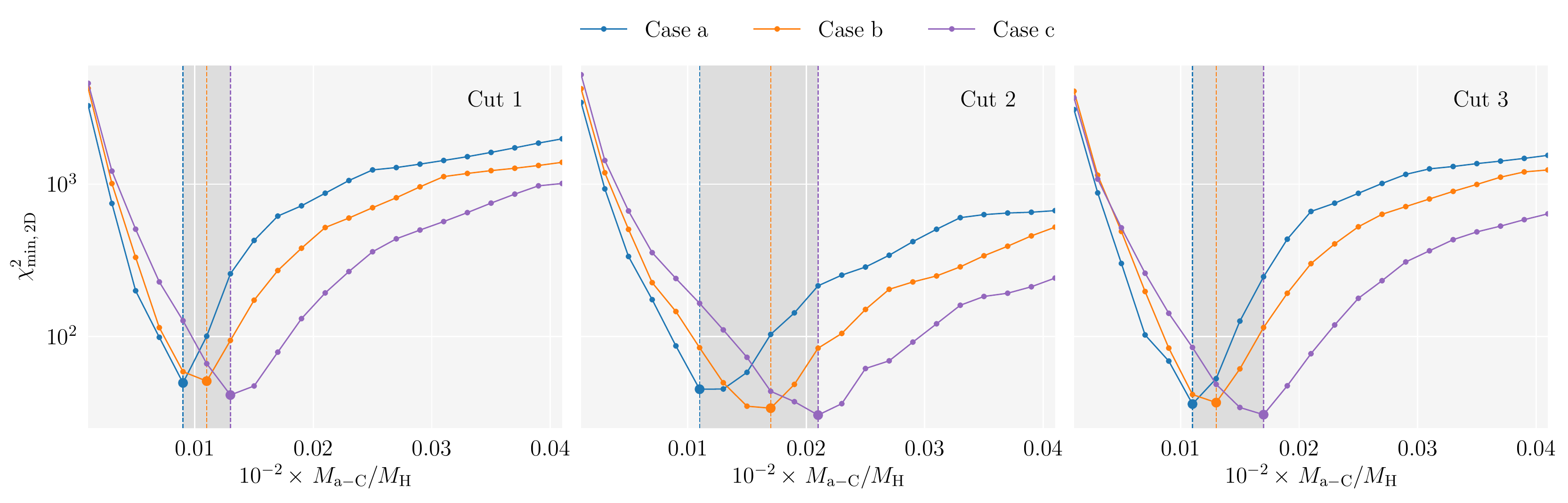}
\caption{\chimindab~as a function of \abvsg~
for cut 1 (left), cut 2 (middle) and cut 3 (right). Blue lines refer to case a,
orange lines to the case b and purple lines to case c. Vertical dashed-lines 
correspond to the minima of \chimind~(i.e. \chimin).}
\label{fig:SOC_chi2min_final}
\end{figure*}

Figure\,\ref{fig:SOC_chi2min_final} shows \chimindab~and Tab.\,\ref{tab:best_fit} 
summarises these results.

First and foremost, \abvsg~is between 0.01 $\times 10^{-2}$ and 0.03 $\times 10^{-2}$, i.e. between 6 to 10 times
lower than in the diffuse ISM (0.17 $\times 10^{-2}$) regardless of
the cut or the case considered. Second, \aminvsg~is between 0.8 and 0.925 nm, i.e.
between 2 and 2.25 times larger than in the diffuse 
ISM (0.4 nm). Third, $\alpha$ is between -7 and -5.5, i.e. between 1.1 
to 1.4 times lower than in the diffuse ISM (-5).

One may note that regardless of the cut, \abvsg~increases from
case \textit{a}
to case \textit{c} (see Fig.\,\ref{fig:SOC_chi2min_final}). In case \textit{a}, we only use modified CM grains (i.e. CM grains with values of \abvsg, \aminvsg~and $\alpha$ that differ from the diffuse ISM)
in the outer part and in the inner part of the Horsehead but we use modified CM-grains in the outer part of the Horsehead and AMMI in the inner part in case \textit{c}.
As AMMI are more emissive in the far-IR than CM grains, emission in this wavelength range
must decrease to fit the observations, which is achievable by reducing \lpdr~(see 
Sect.\,\ref{sec:sub:sub:lpdr_radiative_transfer}), hence \lpdr~decreases from case a 
to case \textit{c}. This decrease in \lpdr~implies a decrease in emission
in the near and mid-IR hence \abvsg~must increase to counterbalance this variation.

\begin{table*}[h]
    \centering
    \begin{tabular}{l|ccc|ccc|ccc}
        \hline
        \hline
        & \multicolumn{3}{c|}{Case a} & \multicolumn{3}{c|}{Case b} & \multicolumn{3}{c}{Case c}\\
         & cut 1 & cut 2 & cut 3 & cut 1 & cut 2 & cut 3 & cut 1 & cut 2 & cut 3 \\ 
         \hline
         $10^{2}$ $\times$ \abvsg & 0.009 & 0.011 & 0.011 & 0.011 & 0.017 & 0.013  & 0.013 & 0.021 & 0.017 \\
         \aminvsg~[nm] & 0.825 & 0.825 & 0.925 & 0.825 & 0.8 & 0.925 & 0.825 & 0.8 & 0.9 \\
         $\alpha$ & -7.0 & -6.0 & -7.5  & -6.5 & -5.5 & -7.5 & -6.5 & -5.5 & -6.5\\
         \lpdr~[pc] & 0.283 & 0.297 & 0.273 & 0.290 & 0.267 & 0.282 & 0.275 & 0.254 & 0.265 \\
         \chimin & 49.6 & 45.1 & 36.0 & 51.0 & 33.9 & 36.9 & 41.3 & 30.5 & 30.7 \\
         \hline
    \end{tabular}
    \caption{Best set of parameters (\abvsg, \aminvsg, $\alpha$ and
    \lpdr) and the \chimin~associated with all cuts
    and cases.}
    \label{tab:best_fit}
\end{table*}

\subsubsection{\chii~distribution in the 2D-spaces (\aminvsg, $\alpha$), (\abvsg, \aminvsg) and (\abvsg, $\alpha$)}
\label{sec:sec:a_min_vsg_alpha_ab_vsg}
\begin{figure*}[h]
\centering
\includegraphics[width=1.0\textwidth, trim={0.0cm 0.0cm 0.0cm 0.0cm},clip]{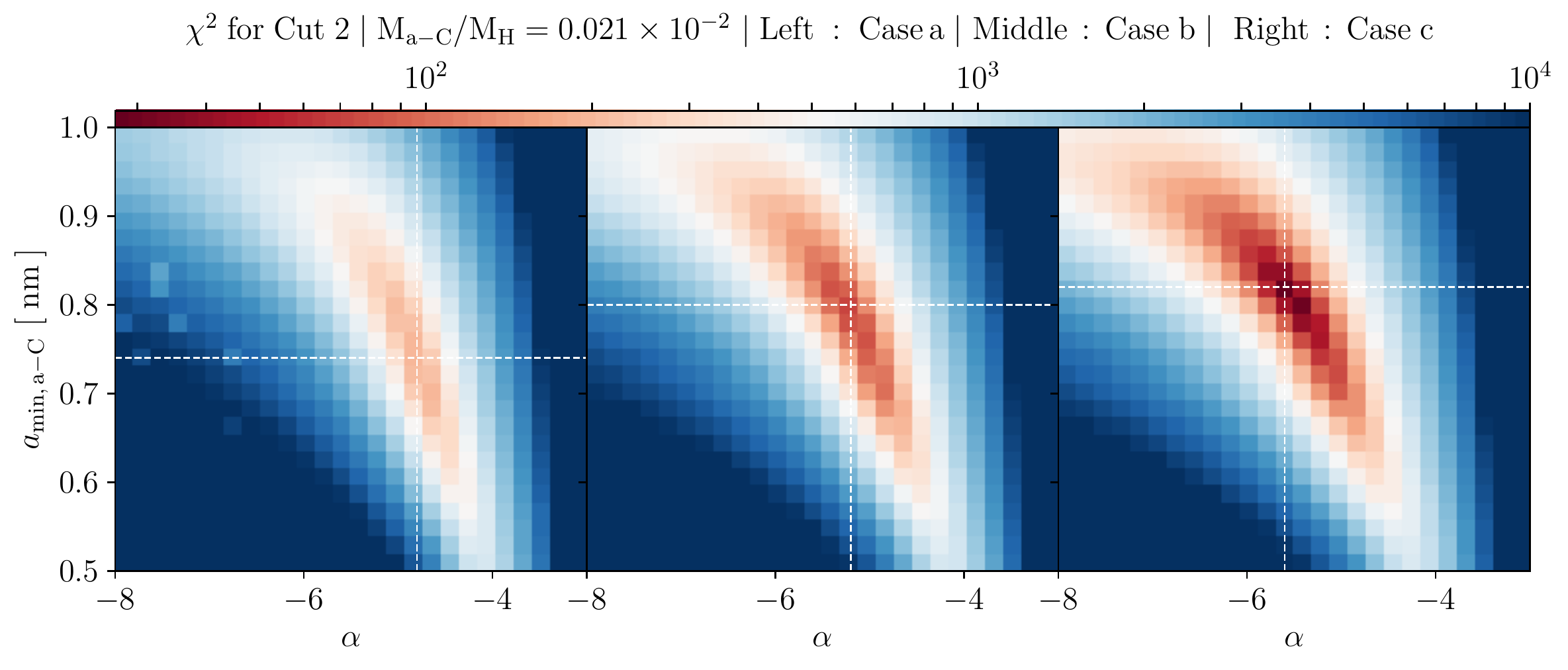}
\includegraphics[width=1.0\textwidth, trim={0.0cm 0.0cm 0.0cm 0.0cm},clip]{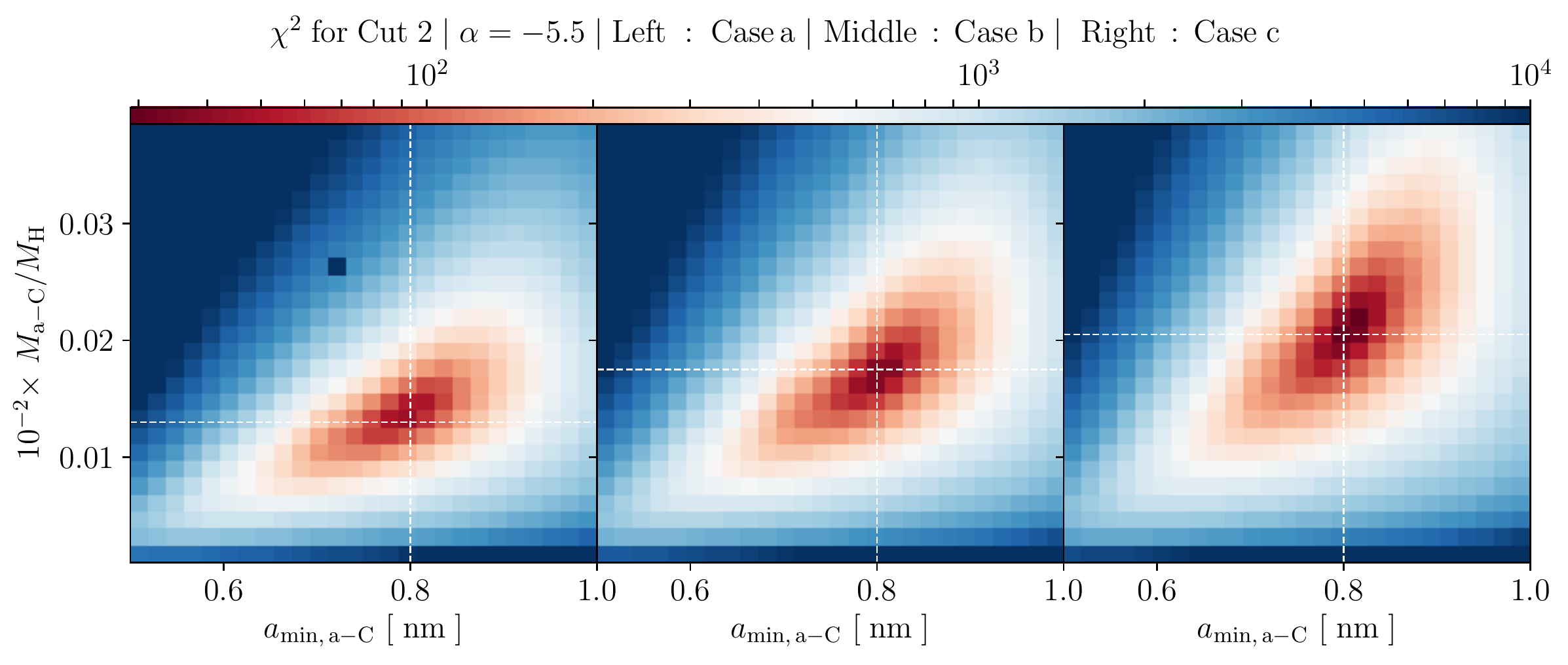}
\includegraphics[width=1.0\textwidth, trim={0.0cm 0.0cm 0.0cm 0.0cm},clip]{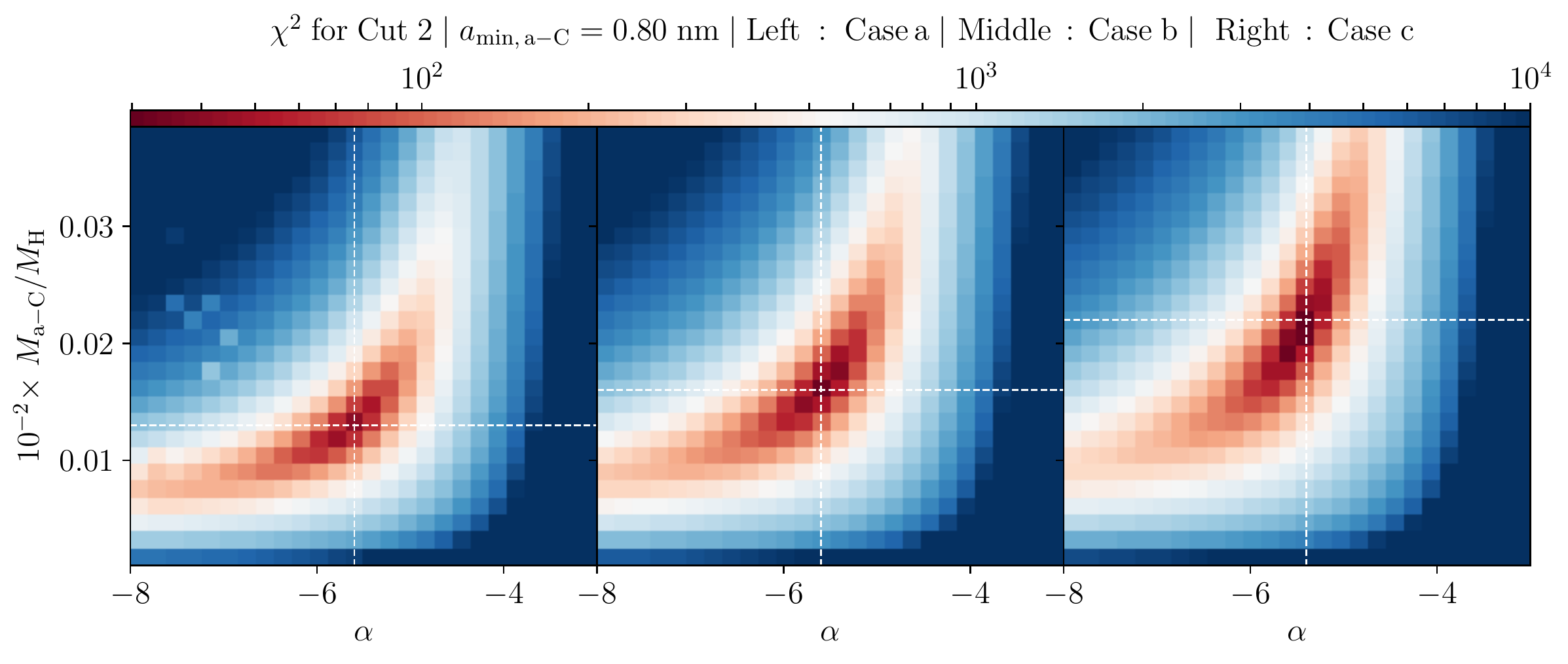}
\caption{Top : \chii~in the 2D-space (\aminvsg, $\alpha$) for 
cut 2 with \abvsg~= 0.021 $\times$ $10^{-2}$. From left to right, subplots correspond 
to case \textit{a} (left), case \textit{b} (middle), case \textit{c} (right). Middle : \chii~in the 2D-space (\abvsg, \aminvsg) for cut 2 with $\alpha$ = -5.5. Bottom : \chii~in the 2D-space (\abvsg, $\alpha$) for cut 2 with \aminvsg~= 0.80 nm.}
\label{fig:SOC_grid_TOT}
\end{figure*}

We show in Fig.\,\ref{fig:SOC_grid_TOT}, the \chii~distribution for cut 2
in the 2D-spaces (\aminvsg, $\alpha$), (\abvsg, \aminvsg) and (\abvsg, $\alpha$).
We choose to focus on only one cut as we are interested 
    in the behaviour of the \chii~distribution here, which is the same
    regardless of the cut. 

The most important result is that, regardless of the case, there is 
a unique minimum in all 2D-spaces. Also, as explained in Sect.\,\ref{sec:sec:DustEM},
a decrease in \abvsg~is, to first order, similar to an increase
in \aminvsg~and $\alpha$, regarding dust emission in the near and mid-
IR. An increase in \aminvsg~is therefore counterbalanced by a 
decrease in $\alpha$ to keep low values of \chii, and an increase in
\abvsg~is counterbalanced by an increase in \aminvsg~and in $\alpha$ hence
explaining the banana-shape of the low \chii~values 
in each of the 2D-spaces. 

It can also be seen that from case \textit{a} to case \textit{c}, the position of 
\chii~minimum value moves. From case \textit{a} to case \textit{c}, dust emission
in the far-IR increases (see Fig.\,\ref{fig:I_emi_CM_AMM_AMMI}) hence
this effect is counterbalanced by a decrease in \lpdr~that also 
reduces dust emission in the near and mid-IR. To compensate for
this decrease in dust emission in the near and mid-IR, the value 
of \abvsg~associated with the \chii~minimum value increases from case
a to case c in the 2D-spaces (\abvsg, \aminvsg) and (\abvsg, $\alpha$).
In the 2D-space (\aminvsg, $\alpha$), this effect is counterbalanced by a decrease
in $\alpha$ and an increase in \aminvsg.

\subsubsection{Comparison between dust modelled emission and dust observed
emission profiles}
\label{sec:sec:dust_emission_profiles}
\begin{figure*}[h]
\centering
\includegraphics[width=1.0\textwidth, trim={0.0cm 0.0cm 0.0cm 0.0cm},clip]{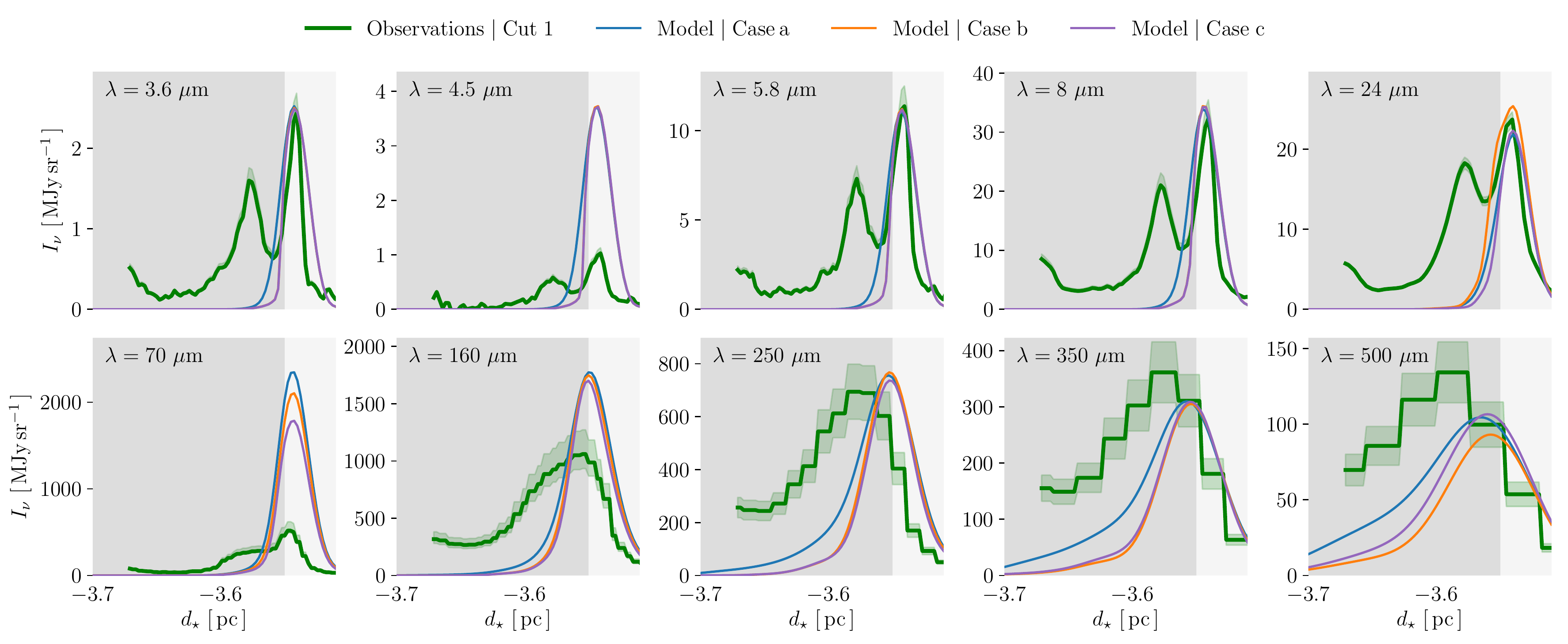}
\includegraphics[width=1.0\textwidth, trim={0.0cm 0.0cm 0.0cm 0.0cm},clip]{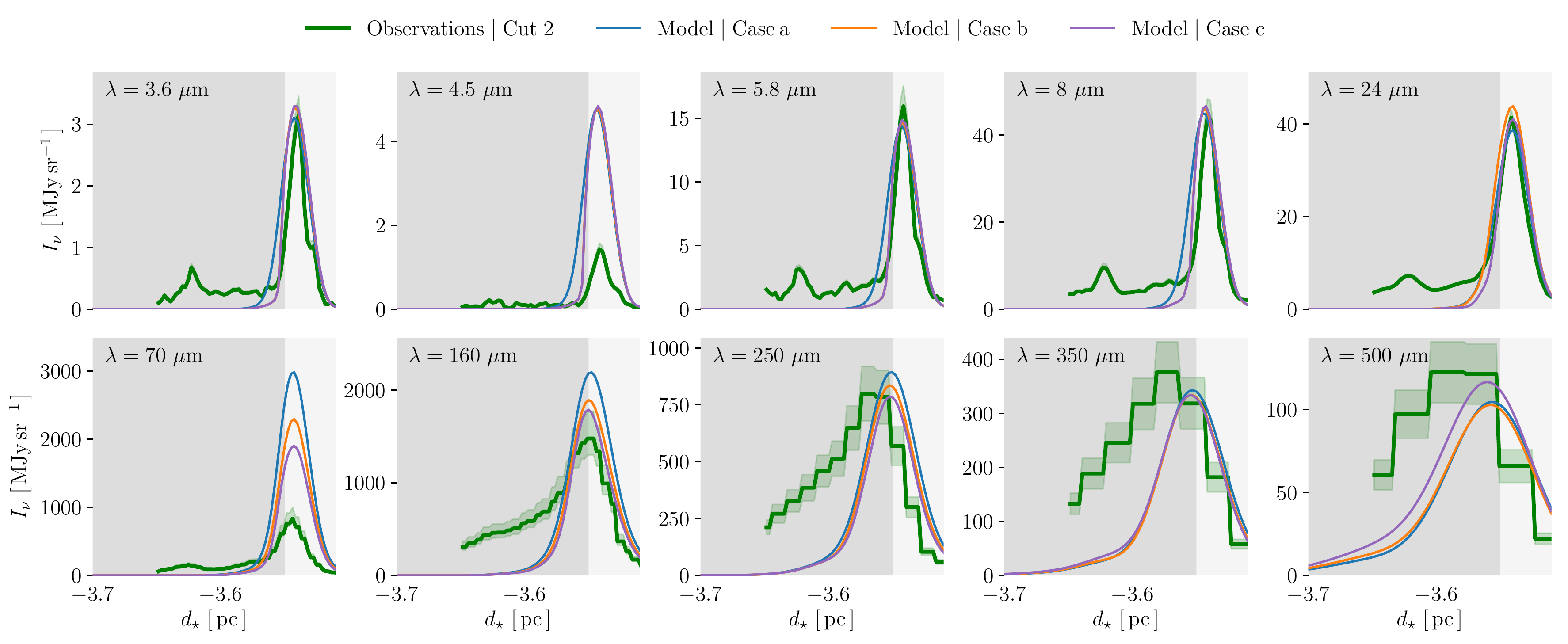}
\includegraphics[width=1.0\textwidth, trim={0.0cm 0.0cm 0.0cm 0.0cm},clip]{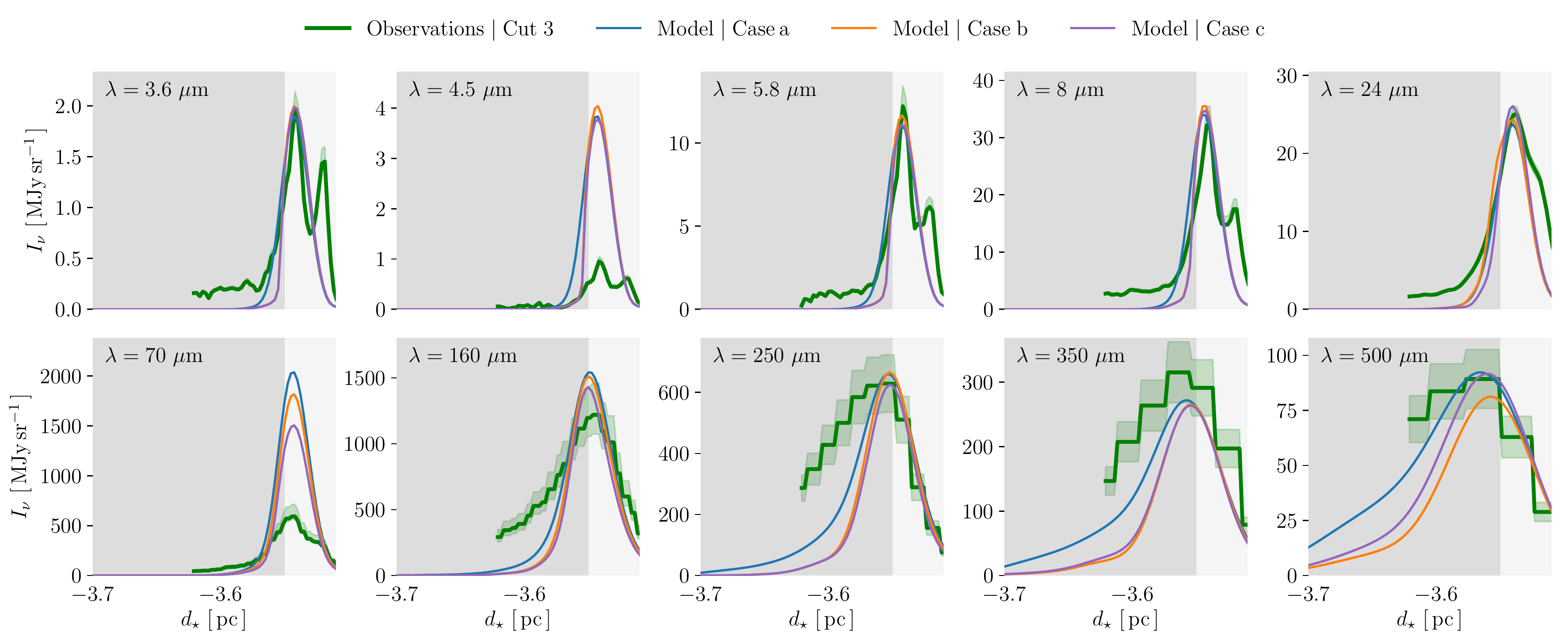}
\caption{Top : Comparison between observed emission profiles for cut 1 
(green line) with modelled emission profiles obtained with the best set of 
parameters (see Tab.\,\ref{tab:best_fit}) for case a (blue line), 
case b (orange line), case c (purple line). Middle : same for cut 2. Bottom : same for
cut 3. The grey parts correspond to the inner
     Horsehead where AMM and AMMI are used in case b and case c,
      respectively.}
\label{fig:SOC_final_obs}
\end{figure*}

Here, we use the best set of parameters (\abvsg, \aminvsg, $\alpha$ and \lpdr) that
are listed in Table.\,\ref{tab:best_fit} and compare the modelled emission profiles
in the 10 photometric bands for the three cases with the observed emission profiles in the three cuts (see Fig.\,\ref{fig:SOC_final_obs}). We focus on three aspects: the maximum of intensity 
in each of the 10 bands, the position of these maxima and the width 
of these profiles.

In the near and mid-IR, except in \IRACdeux, the maximum emission is
well reproduced, regardless the case or the cut. In \MIPSdeux, 
although the maximum of emission is never reproduced, the 
discrepancy between the maximum modelled emission and the 
maximum observed emission decreases from case \textit{a} to case \textit{c}.
From \SPIREun~to \SPIREtrois, the maximum emission is in the error
bars, regardless of the case or the cut and the discrepancy 
between the maximum modelled emission and the maximum observed 
emission decreases from case a to case c. Regarding \MIPStrois, 
the maximum emission is within the error bars only for case \textit{c} 
for cuts 2 and 3 but never for cut 1, 
regardless of the case. 

Concerning the position of the maximum emission, it is well reproduced from \IRACun~
to \MIPSdeux~regardless of the cut and the case. Regarding the cut 1, there is a small
discrepancy between the position of the maximum emission and the position of the observed emission from \MIPSdeux~to \SPIREtrois. For 
cut 2, there is the same discrepancy in \SPIREdeux~and \SPIREtrois, regardless of the case.
For cut 3, all the positions are well reproduced.  

Regarding the width of the profiles, they are well reproduced 
from \IRACun~to \MIPStrois~but slightly different from \SPIREun~to 
\SPIREtrois, which could be due to large structures in the Horsehead. 

To summarise, the observed dust emission is well reproduced in the 
near and mid-IR, except in \IRACdeux, regardless of the case and the 
cut. In the far-IR, the discrepancy between observed dust emission 
and modelled dust emission decreases from case \textit{a} to case \textit{c}.

\section{Discussion}
\label{sec:discussion}
First, we discuss the discrepancy between the dust modelled emission
and the dust observed emission in \IRACdeux~and in \MIPSdeux; second,
the results obtained are described; third, we propose a scenario
of dust evolution in agreement with the results obtained. We end with a
discussion of dust processing timescales in support to this
scenario.

\subsection{Discrepancy in \IRACdeux~and \MIPSdeux}
\label{sec:discrepancy}
In \IRACdeux, the modelled dust emission is always overestimated (see
Fig.\,\ref{fig:SOC_final_obs}) by a factor 2 to 4. As this filter covers
the dust continuum and the wings of the IR bands from a-C:H nano-grains, this suggests that the wings of the IR bands in 
this region are different (i.e., weaker and/or narrower, see for instance 
\cite{bouteraon_carbonaceous_2019} for more details about the variability of 
the IR band widths) from those in the diffuse ISM. Indeed, we are here looking at
dust that is evolving from dense cloud dust in response to 
interaction with UV photons.

Moreover, a-C:H nano-grains freshly produced may not yet have
time to be entirely photo-processed, hence have a large band-to-continuum
ratio because of their high fraction of aliphatic bonds, as opposed to aromatic bonds. As discussed in \cite{jones_evolution_2013},
this requires a-C:H nano-grains with a band gap larger than 0.1 eV, the value adopted in the diffuse ISM.

However, as we do not have dust spectroscopic informations 
of the Horsehead in the near-IR, we are not able to answer to
these previous
questions. In the near future, JWST spectroscopic data should
allow us to understand such changes in the structure of a-C:H
nanograins.

In \MIPSdeux, models always overestimate the emission (see
Fig.\,\ref{fig:SOC_final_obs}) by a factor 3 to 4. This suggests that large grains (\py~and \CM~for 
case \textit{a}, AMM for case \textit{b} and AMMI for case \textit{c}) are somewhat too warm and not emissive 
enough. This is supported by recent laboratory experiments in which the mass
absorption coefficient of silicates in the far-IR is larger (up to an
order of magnitude) than those currently used in THEMIS \citep[see Fig.\,5 in][]{demyk_low_2017}. As a consequence, the large grains we use in there are probably not
emissive enough. The incorporation of these new laboratory results in THEMIS will most likely
reduces the discrepancy in \MIPSdeux.

\subsection{Main results}
Using the 3D radiative transfer code SOC together with the dust
model THEMIS,
we can reproduce the Horsehead observations
in 8 of the 10 photometric bands of \spitzer~and \herschel.

The main results for the outer part of the Horsehead are
the following~:

\begin{enumerate}
    \item The nano-grains (i.e. 
    \cm~grains) dust-to-gas mass ratio, \abvsg, is 6 to 10 times lower than in the diffuse
    ISM.
    \item The minimum size of the nano-grains, 
    \aminvsg, is 2 to 2.25 times larger than in the diffuse ISM.
    \item The power-law exponent of the nano-grains size distribution, $\alpha$, is 1.1 to 1.4 times lower than in the diffuse ISM, i.e. the size distribution is steeper. 
\end{enumerate}

The best size distributions for the three cuts and
case \textit{c} are shown in Fig.\,\ref{fig:s_dist_final}. 
Concerning the inner part of the Horsehead, we tested 3 different kinds
of dust, diffuse ISM-like dust (CM) with modified size distributions in case \textit{a}, aggregates 
of grains (AMM) in case \textit{b}, aggregates of grains with ice mantles
(AMMI) in case \textit{c}. At long wavelengths (from \MIPStrois~to \SPIREtrois) The results are
significantly better when using AMMI instead of CM grains.
Regarding \MIPSdeux, even if we are not able to reproduce the observed
emission with our model, using aggregates (AMM/AMMI) instead of diffuse ISM-like dust (CM) with modified size distributions, significantly improves the fit in this band.  

Finally, the length of the Horsehead along the line-of-sight, \lpdr, 
is found to be within the range of 0.26 and 0.30 pc which is in agreement with 
previous gas studies \citep{habart_density_2005}. 

\subsection{Dust evolution scenario}
\label{sub:sub:scenario}

Our results show significant variations of the dust size distribution and in the following we outline a possible
scenario of dust evolution across the Horsehead interface. Given the strong incident radiation field, we assume that the dominant process is the exposure 
of dust grains from the dense molecular cloud (the inner region) to the UV light of $\sigma$-Ori.
This suggests two major photo-processing sequences: (i) the partial fragmentation of aggregate grains from the inner region
and (ii) the destruction of the smallest a-C:H nano-grains. We discuss the significance of these sequences by comparing
their timescales to the advection timescale $\tau_{a}$, i.e., the time that the incident UV light needs to heat up and
dissociate the molecular gas at the cloud border. 

The advection timescale is defined as $\tau_{\mathrm{a}}=L/v_{\mathrm{DF}}$ where 
$L\sim 0.05$ pc is the width of the outer part of the Horsehead and 
$v_{\mathrm{DF}}\sim 0.5$ km.s$^{-1}$ is the velocity of the dissociation front 
\citep{hollenbach_photodissociation_1999}. With these values, we find $\tau_{\mathrm{a}}\sim 10^{5}$ years.

Due to the lack of
studies, we take the photo-darkening timescale as a lower limit to photo-fragmentation of aggregate grains and
 photo-destruction of \cm~nano-grains, described by $\tau_{\mathrm{ph}}$.
Indeed, photo-darkening involves the dissociation of CH-bonds, a process that is more likely faster than the breaking of CC-bonds
that must occur in photo-fragmentation \citep{jones_h_2015}. We thus express $\tau_{\mathrm{ph}}$ at the cloud edge in terms of the photo-darkening rate $\Lambda_{\mathrm{pd}}$ \citep{jones_cycling_2014} :
\begin{equation}
\tau_{\mathrm{ph}} \simeq \Lambda_{\mathrm{pd}}^{-1} = {1\over {\sigma_{\mathrm{CH}}\,F^0_{\mathrm{UV}}\,Q_{\mathrm{abs}}(a)\,\epsilon(a)}},
\end{equation}
where $F^0_{\mathrm{UV}}\simeq 3.8\times 10^9$ photons.s$^{-1}$.cm$^{-2}$ is the unattenuated UV field, $\epsilon(a)={\rm min}(1,{2\over a[{\rm nm}]})$ is a size-dependent 
photo-darkening efficiency, $\sigma_{\mathrm{CH}}\simeq 10^{-19}$ cm$^{2}$
is the CH bond photo-dissociation cross-section and
$Q_{\mathrm{abs}}(a)$,
the dust absorption efficiency which depends almost solely on the 
radius in the UV range. In the case of AMMI, $\tau_{\mathrm{ph}}$ is larger in reality because 
the ice mantle needs to be vaporised first but we do not take into account this effect as the time $\tau_{\mathrm{ph}}$
we estimate is already a lower limit. 

\begin{figure}[h]
\centering
\includegraphics[width=0.5\textwidth, trim={0.0cm 0.0cm 0.0cm 0.0cm},clip]{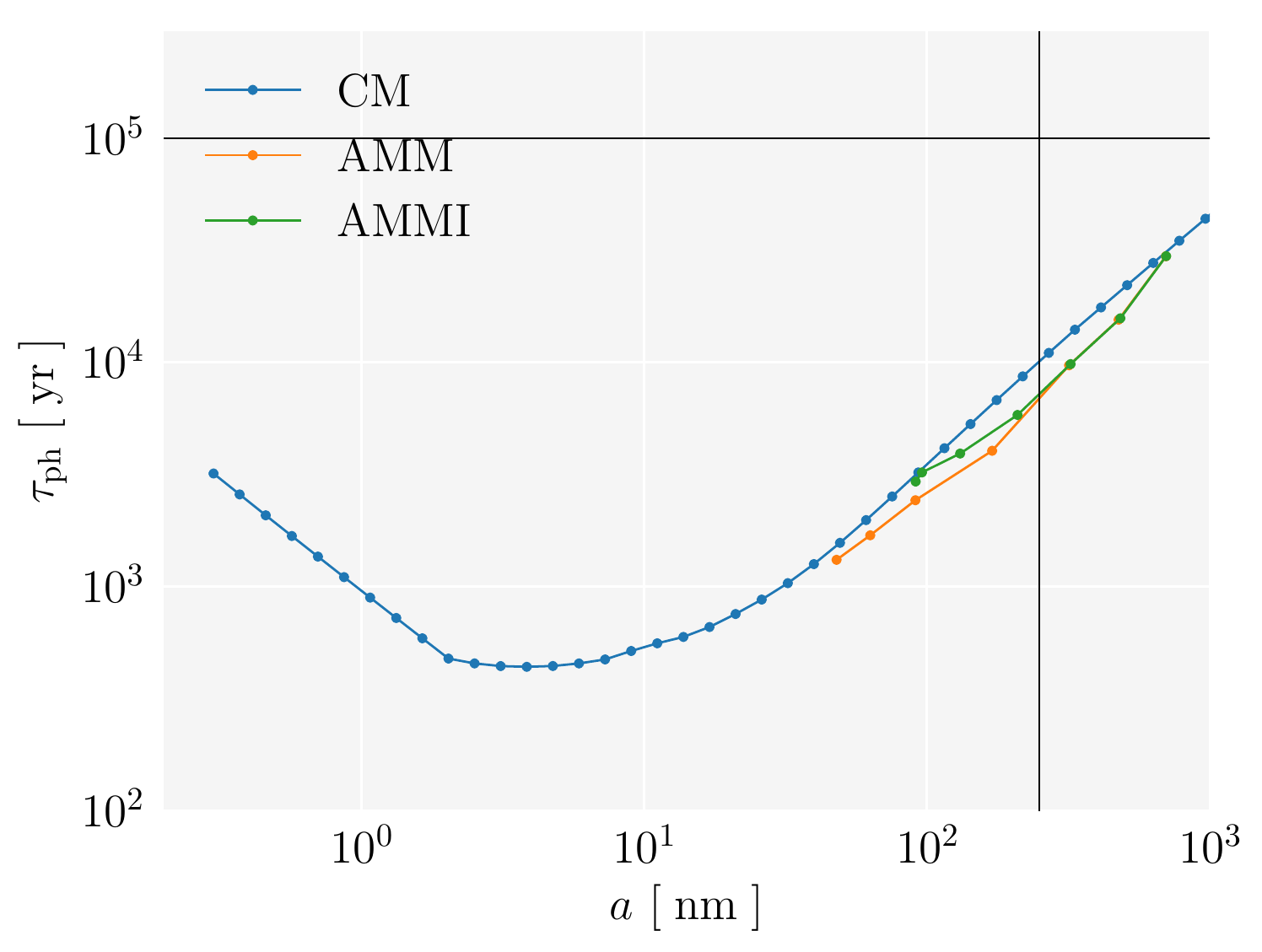}
\caption{Photo-fragmentation time-scale at the Horsehead edge as a function of the
grain radius, $a$. Blue line refers to CM grains, orange line to AMM and green 
line to AMMI. The horizontal black line corresponds to the advection time-scale 
$\tau_{\mathrm{a}}\sim 10^{5}$ years in the outer part of the Horsehead. 
The vertical black line corresponds to the limit at 250 nm 
beyond which more than 50~$\%$ of the AMM(I) dust mass is contained \citep{ysard_mantle_2016}.}
\label{fig:photodarkening}
\end{figure}

We show $\tau_{\mathrm{ph}}(a)$ in Fig.\,\ref{fig:photodarkening} for CM, 
AMM and AMMI. As discussed in \cite{ysard_mantle_2016}, more than 50~$\%$ of the AMM(I)
dust mass is contained in grains larger than 250 nm. From this figure, one can see that 
aggregate grains can be photo-fragmented because $\tau_{\mathrm{ph}}\sim \tau_{\mathrm{a}}$. One can also see that \cm~nano-grains can be 
efficiently destroyed as $\tau_{\mathrm{ph}}<\tau_{\mathrm{a}}$ 
for \cm~nano-grains. Similar results were found by \cite{alata_vacuum_2014}, from laboratory experiments on a-C:H grain analogues, later applied to the Horsehead \citep{alata_vacuum_2015}.

\begin{figure}[h]
\centering
\includegraphics[width=0.5\textwidth, trim={0.0cm 0.0cm 0.0cm 0.0cm},clip]{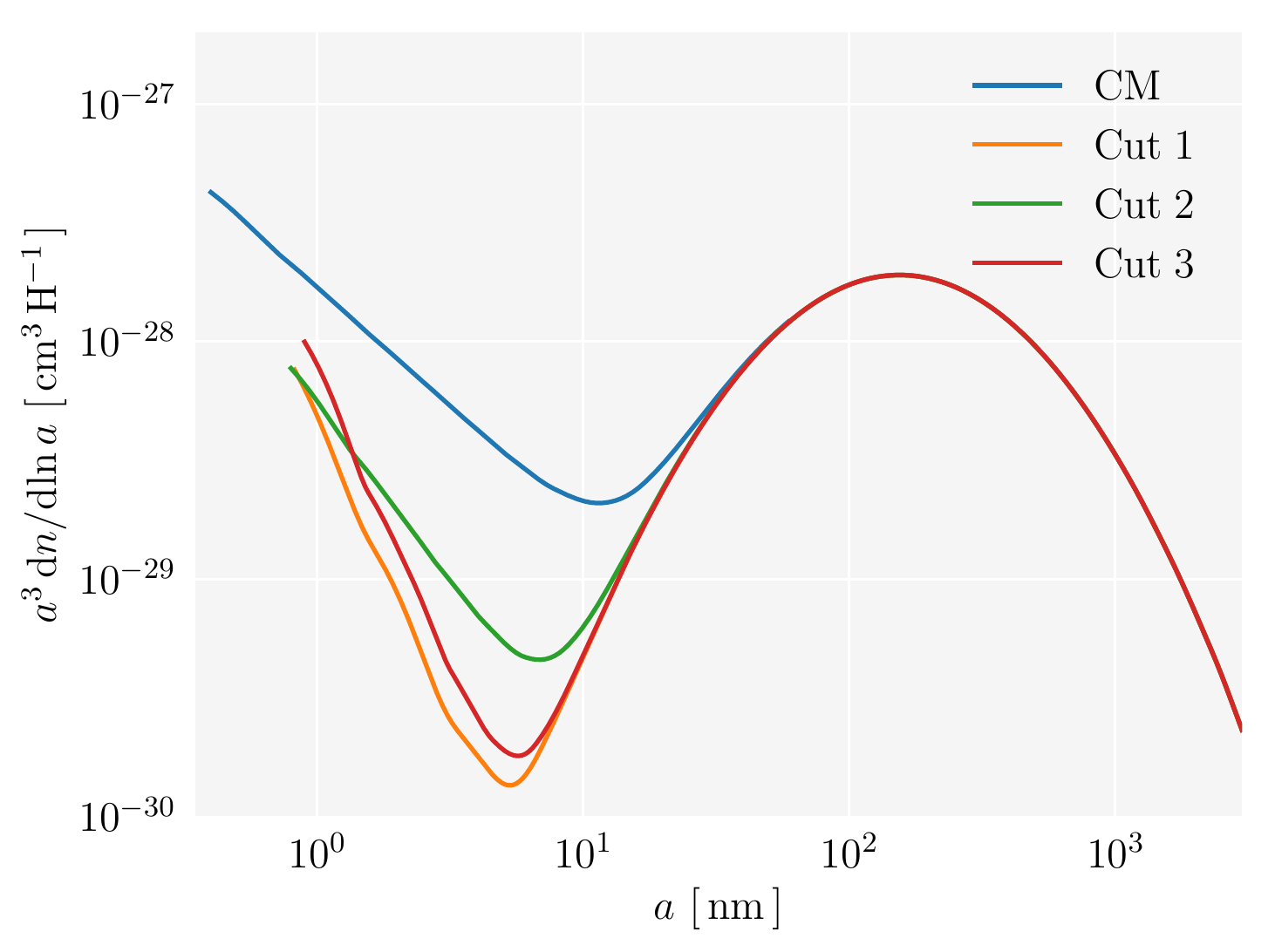}
\caption{Size distribution for diffuse ISM-like dust (CM) in blue. 
Modified size distributions using the best set of parameters (see Tab.\,\ref{tab:best_fit}) in case \textit{c} for cut 1 (orange), cut 2 (green), cut 3 (red).}
\label{fig:s_dist_final}
\end{figure}

From this analysis emerges the following scenario.
Within an advection timescale, the a-C nano-grains formed by fragmentation of aggregate grains are also partially 
destroyed by UV photons. This naturally explains the depletion of a-C:H grains around a=10 nm seen in Fig.\,\ref{fig:s_dist_final}). We
note that the size distribution of these freshly formed small grains is significanlty different from the diffuse ISM case (blue curve in Fig.\,\ref{fig:s_dist_final})). This evolved size distribution could reflect the photo-evaporated layer described by \cite{bron_photoevaporating_2018}.

\section{Conclusion}
\label{sec:conclusion}
With \herschel~and \spitzer~data, we studied the Horsehead
using 10 photometric bands, from 3.6
\mum~to 500 \mum, covering the
entire dust spectrum. We modelled the dust emission across the Horsehead
using the THEMIS
dust model together with the 3D radiative transfer code SOC. 

We show that it is not possible to reproduce the observations in
the Horsehead using dust grains from the diffuse ISM hence the necessity to
modify their size distributions and compositions. Dust therefore evolves across
the Horsehead. 

In the outer part of the Horsehead, the \cm~nano-grains dust-to-gas ratio is 6 to 10 times lower
and their minimum size, 2
to 2.25 times larger than in the diffuse ISM. The power law 
of the size distribution is steeper than in the diffuse ISM. In the inner
part of the Horsehead, we show that using aggregate grains 
with or without ice mantles significantly reduces the discrepancy between
our model and the observations. The discrepancy between the observations and our model at 4.5 \mum~could
be due to the shape of the aromatic band wings whence the overestimation
of the dust modelled emission. We also find that large grains are too warm because our modelled dust emission at 70 \mum~is overestimated. However, laboratory
studies show that large silicate grains 
are more emissive than those used in dust models hence cooler. These new
results will soon be implemented in THEMIS.

Based on a time-scale analysis, we propose a scenario where the
\cm~nano-grains form by the partial
photo-fragmentation of aggregate grains and are processed by the 
UV photons, leading to a size distribution depleted in grains of size 
from 5 to 10 nm. In the denser regions of the Horsehead, the dust composition is typical
of dense clouds.

Spectroscopic observations of the Horsehead are required to go further on the structure and size distribution of \cm~nano-grains. Indeed, the
observations with the JWST will, for the first time, spatially resolve the
individual IR dust signatures across the Horsehead, offering an unprecedented
look at the evolution of the interstellar matter in photon-dominated regions.

\begin{acknowledgements} 
We would like to thank the CNES and the P2IO LabeX for supporting Thiebaut Schirmer PhD work.

This work was supported by the Programme National “Physique et
Chimie du Milieu Interstellaire” (PCMI) of CNRS/INSU with INC/INP co-funded
by CEA and CNES.
HIPE is a joint development by the Herschel Science Ground Segment Consortium, consisting of ESA, the NASA Herschel Science Center, and the HIFI, PACS and SPIRE consortia. PACS has been developed by a consortium of institutes led by MPE (Germany) and including UVIE (Austria); KU Leuven, CSL, IMEC (Belgium); CEA, LAM (France); MPIA (Germany); INAF-IFSI/OAA/OAP/OAT, LENS, SISSA (Italy); IAC (Spain). This development has been supported by the funding agencies BMVIT (Austria), ESA-PRODEX (Belgium), CEA/CNES (France), DLR (Germany), ASI/INAF (Italy), and CICYT/MCYT (Spain).
SPIRE has been developed by a consortium of institutes led by Cardiff Univ. (UK) and
including Univ. Lethbridge (Canada); NAOC (China); CEA, LAM (France); IFSI, Univ.
Padua (Italy); IAC (Spain); Stockholm Observatory (Sweden); Imperial College London,
RAL, UCL-MSSL, UKATC, Univ. Sussex (UK); Caltech, JPL, NHSC, Univ. Colorado (USA).
This development has been supported by national funding agencies: CSA (Canada); NAOC
(China); CEA, CNES, CNRS (France); ASI (Italy); MCINN (Spain); SNSB (Sweden); STFC,
UKSA (UK); and NASA (USA)
\end{acknowledgements}

\bibliography{Zotero}

\begin{thebibliography}{108}
\expandafter\ifx\csname natexlab\endcsname\relax\def\natexlab#1{#1}\fi

\bibitem[{Abergel {et~al.}(2010)Abergel, Arab, Compiègne, Kirk, Ade, Anderson,
  André, Baluteau, Bernard, Blagrave, Bontemps, Boulanger, Cohen, Cox,
  Dartois, Davis, Emery, Fulton, Gry, Habart, Huang, Joblin, Jones, Lagache,
  Lim, Madden, Makiwa, Martin, Miville-Deschênes, Molinari, Moseley, Motte,
  Naylor, Okumura, Pinheiro~Gonçalves, Polehampton, Rodon, Russeil, Saraceno,
  Sauvage, Sidher, Spencer, Swinyard, Ward-Thompson, White, \&
  Zavagno}]{abergel_evolution_2010}
Abergel, A., Arab, H., Compiègne, M., {et~al.} 2010, Astronomy and
  Astrophysics, 518, L96

\bibitem[{Abergel {et~al.}(1994)Abergel, Boulanger, Mizuno, \&
  Fukui}]{abergel_comparative_1994}
Abergel, A., Boulanger, F., Mizuno, A., \& Fukui, Y. 1994, The Astrophysical
  Journal Letters, 423, L59

\bibitem[{Abergel {et~al.}(2003)Abergel, Teyssier, Bernard, Boulanger, Coulais,
  Fosse, Falgarone, Gerin, Perault, Puget, Nordh, Olofsson, Huldtgren, Kaas,
  André, Bontemps, Casali, Cesarsky, Copet, Davies, Montmerle, Persi, \&
  Sibille}]{abergel_isocam_2003}
Abergel, A., Teyssier, D., Bernard, J.~P., {et~al.} 2003, Astronomy and
  Astrophysics, 410, 577

\bibitem[{Alata {et~al.}(2014)Alata, Cruz-Diaz, Muñoz~Caro, \&
  Dartois}]{alata_vacuum_2014}
Alata, I., Cruz-Diaz, G.~A., Muñoz~Caro, G.~M., \& Dartois, E. 2014, Astronomy
  and Astrophysics, 569, A119

\bibitem[{Alata {et~al.}(2015)Alata, Jallat, Gavilan, Chabot, Cruz-Diaz,
  Munoz~Caro, Béroff, \& Dartois}]{alata_vacuum_2015}
Alata, I., Jallat, A., Gavilan, L., {et~al.} 2015, Astronomy and Astrophysics,
  584, A123

\bibitem[{Anthony-Twarog(1982)}]{anthony-twarog_h-beta_1982}
Anthony-Twarog, B.~J. 1982, The Astronomical Journal, 87, 1213

\bibitem[{Arab(2012)}]{arab_evolution_2012-2}
Arab, H. 2012, PhD thesis

\bibitem[{Arab {et~al.}(2012)Arab, Abergel, Habart, Bernard-Salas, Ayasso,
  Dassas, Martin, \& White}]{arab_evolution_2012}
Arab, H., Abergel, A., Habart, E., {et~al.} 2012, Astronomy and Astrophysics,
  541, A19

\bibitem[{Bakes \& Tielens(1994)}]{bakes_photoelectric_1994}
Bakes, E. L.~O. \& Tielens, A. G. G.~M. 1994, The Astrophysical Journal, 427,
  822

\bibitem[{Bernard {et~al.}(1999)Bernard, Abergel, Ristorcelli, Pajot, Torre,
  Boulanger, Giard, Lagache, Serra, Lamarre, Puget, Lepeintre, \&
  Cambrésy}]{bernard_pronaos_1999}
Bernard, J.~P., Abergel, A., Ristorcelli, I., {et~al.} 1999, Astronomy and
  Astrophysics, 347, 640

\bibitem[{Berné {et~al.}(2007)Berné, Joblin, Deville, Smith, Rapacioli,
  Bernard, Thomas, Reach, \& Abergel}]{berne_analysis_2007}
Berné, O., Joblin, C., Deville, Y., {et~al.} 2007, Astronomy and Astrophysics,
  469, 575

\bibitem[{Boersma {et~al.}(2014)Boersma, Bregman, \&
  Allamandola}]{boersma_properties_2014}
Boersma, C., Bregman, J., \& Allamandola, L.~J. 2014, The Astrophysical
  Journal, 795, 110

\bibitem[{Boulanger {et~al.}(1990)Boulanger, Falgarone, Puget, \&
  Helou}]{boulanger_variations_1990}
Boulanger, F., Falgarone, E., Puget, J.~L., \& Helou, G. 1990, The
  Astrophysical Journal, 364, 136

\bibitem[{Boutéraon {et~al.}(2019)Boutéraon, Habart, Ysard, Jones, Dartois,
  \& Pino}]{bouteraon_carbonaceous_2019}
Boutéraon, T., Habart, E., Ysard, N., {et~al.} 2019, Astronomy \&
  Astrophysics, 623, A135

\bibitem[{Bowler {et~al.}(2009)Bowler, Waller, Megeath, Patten, \&
  Tamura}]{bowler_infrared_2009}
Bowler, B.~P., Waller, W.~H., Megeath, S.~T., Patten, B.~M., \& Tamura, M.
  2009, The Astronomical Journal, 137, 3685

\bibitem[{Bron {et~al.}(2018)Bron, Agúndez, Goicoechea, \&
  Cernicharo}]{bron_photoevaporating_2018}
Bron, E., Agúndez, M., Goicoechea, J.~R., \& Cernicharo, J. 2018,
  arXiv:1801.01547 [astro-ph], arXiv: 1801.01547

\bibitem[{Bron {et~al.}(2014)Bron, Le~Bourlot, \& Le~Petit}]{bron_surface_2014}
Bron, E., Le~Bourlot, J., \& Le~Petit, F. 2014, Astronomy and Astrophysics,
  569, A100

\bibitem[{Burke \& Hollenbach(1983)}]{burke_gas-grain_1983}
Burke, J.~R. \& Hollenbach, D.~J. 1983, The Astrophysical Journal, 265, 223

\bibitem[{Campeggio {et~al.}(2007)Campeggio, Strafella, Maiolo, Elia, \&
  Aiello}]{campeggio_total_2007}
Campeggio, L., Strafella, F., Maiolo, B., Elia, D., \& Aiello, S. 2007, The
  Astrophysical Journal, 668, 316

\bibitem[{Cardelli \& Clayton(1991)}]{cardelli_absolute_1991}
Cardelli, J.~A. \& Clayton, G.~C. 1991, The Astronomical Journal, 101, 1021

\bibitem[{Cardelli {et~al.}(1989)Cardelli, Clayton, \&
  Mathis}]{cardelli_relationship_1989}
Cardelli, J.~A., Clayton, G.~C., \& Mathis, J.~S. 1989, The Astrophysical
  Journal, 345, 245

\bibitem[{Compiègne {et~al.}(2008)Compiègne, Abergel, Verstraete, \&
  Habart}]{compiegne_dust_2008}
Compiègne, M., Abergel, A., Verstraete, L., \& Habart, E. 2008, Astronomy and
  Astrophysics, 491, 797

\bibitem[{Compiègne {et~al.}(2007)Compiègne, Abergel, Verstraete, Reach,
  Habart, Smith, Boulanger, \& Joblin}]{compiegne_aromatic_2007}
Compiègne, M., Abergel, A., Verstraete, L., {et~al.} 2007, Astronomy and
  Astrophysics, 471, 205

\bibitem[{Compiègne {et~al.}(2011)Compiègne, Verstraete, Jones, Bernard,
  Boulanger, Flagey, Le~Bourlot, Paradis, \& Ysard}]{compiegne_global_2011}
Compiègne, M., Verstraete, L., Jones, A., {et~al.} 2011, Astronomy and
  Astrophysics, 525, A103

\bibitem[{de~Boer(1983)}]{de_boer_diffuse_1983}
de~Boer, K.~S. 1983, Astronomy and Astrophysics, 125, 258

\bibitem[{Demyk {et~al.}(2017)Demyk, Meny, Lu, Papatheodorou, Toplis, Leroux,
  Depecker, Brubach, Roy, Nayral, Ojo, Delpech, Paradis, \&
  Gromov}]{demyk_low_2017}
Demyk, K., Meny, C., Lu, X.-H., {et~al.} 2017, Astronomy \& Astrophysics, 600,
  A123

\bibitem[{Desert {et~al.}(1990)Desert, Boulanger, \&
  Puget}]{desert_interstellar_1990}
Desert, F.-X., Boulanger, F., \& Puget, J.~L. 1990, Astronomy and Astrophysics,
  237, 215

\bibitem[{Draine(2003)}]{draine_interstellar_2003}
Draine, B. 2003, Annual Review of Astronomy and Astrophysics, 41, 241

\bibitem[{Draine \& Lee(1984)}]{draine_optical_1984}
Draine, B.~T. \& Lee, H.~M. 1984, The Astrophysical Journal, 285, 89

\bibitem[{Duley {et~al.}(2015)Duley, Zaidi, Wesolowski, \&
  Kuzmin}]{duley_small_2015}
Duley, W.~W., Zaidi, A., Wesolowski, M.~J., \& Kuzmin, S. 2015, Monthly Notices
  of the Royal Astronomical Society, 447, 1242

\bibitem[{Engelbracht {et~al.}(2007)Engelbracht, Blaylock, Su, Rho, Rieke,
  Muzerolle, Padgett, Hines, Gordon, Fadda, Noriega‐Crespo, Kelly, Latter,
  Hinz, Misselt, Morrison, Stansberry, Shupe, Stolovy, Wheaton, Young,
  Neugebauer, Wachter, Pérez‐González, Frayer, \&
  Marleau}]{engelbracht_absolute_2007}
Engelbracht, C., Blaylock, M., Su, K., {et~al.} 2007, Publications of the
  Astronomical Society of the Pacific, 119, 994

\bibitem[{Fitzpatrick \& Massa(1986)}]{fitzpatrick_analysis_1986}
Fitzpatrick, E.~L. \& Massa, D. 1986, The Astrophysical Journal, 307, 286

\bibitem[{Flagey {et~al.}(2009)Flagey, Noriega-Crespo, Boulanger, Carey,
  Brooke, Falgarone, Huard, McCabe, Miville-Deschênes, Padgett, Paladini, \&
  Rebull}]{flagey_evidence_2009}
Flagey, N., Noriega-Crespo, A., Boulanger, F., {et~al.} 2009, The Astrophysical
  Journal, 701, 1450

\bibitem[{Gerin {et~al.}(2009)Gerin, Goicoechea, Pety, \&
  Hily-Blant}]{gerin_hco_2009}
Gerin, M., Goicoechea, J.~R., Pety, J., \& Hily-Blant, P. 2009, Astronomy and
  Astrophysics, 494, 977

\bibitem[{Goicoechea {et~al.}(2009)Goicoechea, Pety, Gerin, Hily-Blant, \&
  Le~Bourlot}]{goicoechea_ionization_2009}
Goicoechea, J.~R., Pety, J., Gerin, M., Hily-Blant, P., \& Le~Bourlot, J. 2009,
  Astronomy and Astrophysics, 498, 771

\bibitem[{Goicoechea {et~al.}(2006)Goicoechea, Pety, Gerin, Teyssier, Roueff,
  Hily-Blant, \& Baek}]{goicoechea_low_2006}
Goicoechea, J.~R., Pety, J., Gerin, M., {et~al.} 2006, Astronomy and
  Astrophysics, 456, 565

\bibitem[{Gordon {et~al.}(2007)Gordon, Engelbracht, Fadda, Stansberry, Wachter,
  Frayer, Rieke, Noriega‐Crespo, Latter, Young, Neugebauer, Balog, Beeman,
  Dole, Egami, Haller, Hines, Kelly, Marleau, Misselt, Morrison,
  Pérez‐González, Rho, \& Wheaton}]{gordon_absolute_2007}
Gordon, K., Engelbracht, C., Fadda, D., {et~al.} 2007, Publications of the
  Astronomical Society of the Pacific, 119, 1019

\bibitem[{Gordon {et~al.}(2017)Gordon, Baes, Bianchi, Camps, Juvela, Kuiper,
  Lunttila, Misselt, Natale, Robitaille, \& Steinacker}]{gordon_trust._2017}
Gordon, K.~D., Baes, M., Bianchi, S., {et~al.} 2017, Astronomy and
  Astrophysics, 603, A114

\bibitem[{Gratier {et~al.}(2013)Gratier, Pety, Guzmán, Gerin, Goicoechea,
  Roueff, \& Faure}]{gratier_iram-30_2013}
Gratier, P., Pety, J., Guzmán, V., {et~al.} 2013, Astronomy and Astrophysics,
  557, A101

\bibitem[{Guzmán {et~al.}(2011)Guzmán, Pety, Goicoechea, Gerin, \&
  Roueff}]{guzman_h2co_2011}
Guzmán, V., Pety, J., Goicoechea, J.~R., Gerin, M., \& Roueff, E. 2011,
  Astronomy and Astrophysics, 534, A49

\bibitem[{Guzmán {et~al.}(2012)Guzmán, Pety, Gratier, Goicoechea, Gerin,
  Roueff, \& Teyssier}]{guzman_iram-30m_2012}
Guzmán, V., Pety, J., Gratier, P., {et~al.} 2012, Astronomy and Astrophysics,
  543, L1

\bibitem[{Guzmán {et~al.}(2013)Guzmán, Goicoechea, Pety, Gratier, Gerin,
  Roueff, Le~Petit, Le~Bourlot, \& Faure}]{guzman_iram-30_2013}
Guzmán, V.~V., Goicoechea, J.~R., Pety, J., {et~al.} 2013, Astronomy and
  Astrophysics, 560, A73

\bibitem[{Habart {et~al.}(2005)Habart, Abergel, Walmsley, Teyssier, \&
  Pety}]{habart_density_2005}
Habart, E., Abergel, A., Walmsley, C.~M., Teyssier, D., \& Pety, J. 2005,
  Astronomy and Astrophysics, 437, 177

\bibitem[{Hily-Blant {et~al.}(2005)Hily-Blant, Teyssier, Philipp, \&
  Güsten}]{hily-blant_velocity_2005}
Hily-Blant, P., Teyssier, D., Philipp, S., \& Güsten, R. 2005, Astronomy and
  Astrophysics, 440, 909

\bibitem[{Hollenbach \& Salpeter(1971)}]{hollenbach_surface_1971}
Hollenbach, D. \& Salpeter, E.~E. 1971, The Astrophysical Journal, 163, 155

\bibitem[{Hollenbach \& Tielens(1997)}]{hollenbach_dense_1997}
Hollenbach, D.~J. \& Tielens, A. G. G.~M. 1997, Annual Review of Astronomy and
  Astrophysics, 35, 179

\bibitem[{Hollenbach \& Tielens(1999)}]{hollenbach_photodissociation_1999}
Hollenbach, D.~J. \& Tielens, A. G. G.~M. 1999, Reviews of Modern Physics, 71,
  173

\bibitem[{Jones(2012{\natexlab{a}})}]{jones_variations_2012-2}
Jones, A.~P. 2012{\natexlab{a}}, Astronomy and Astrophysics, 540, A1

\bibitem[{Jones(2012{\natexlab{b}})}]{jones_variations_2012-1}
Jones, A.~P. 2012{\natexlab{b}}, Astronomy and Astrophysics, 540, A2

\bibitem[{Jones(2012{\natexlab{c}})}]{jones_variations_2012}
Jones, A.~P. 2012{\natexlab{c}}, Astronomy and Astrophysics, 542, A98

\bibitem[{Jones {et~al.}(1990)Jones, Duley, \& Williams}]{jones_structure_1990}
Jones, A.~P., Duley, W.~W., \& Williams, D.~A. 1990, Quarterly Journal of the
  Royal Astronomical Society, 31, 567

\bibitem[{Jones {et~al.}(2013)Jones, Fanciullo, Köhler, Verstraete, Guillet,
  Bocchio, \& Ysard}]{jones_evolution_2013}
Jones, A.~P., Fanciullo, L., Köhler, M., {et~al.} 2013, Astronomy and
  Astrophysics, 558, A62

\bibitem[{Jones \& Habart(2015)}]{jones_h_2015}
Jones, A.~P. \& Habart, E. 2015, Astronomy \& Astrophysics, 581, A92

\bibitem[{Jones {et~al.}(2017)Jones, Köhler, Ysard, Bocchio, \&
  Verstraete}]{jones_global_2017}
Jones, A.~P., Köhler, M., Ysard, N., Bocchio, M., \& Verstraete, L. 2017,
  Astronomy and Astrophysics, 602, A46

\bibitem[{Jones {et~al.}(2014)Jones, Ysard, Koehler, Fanciullo, Bocchio,
  Micelotta, Verstraete, \& Guillet}]{jones_cycling_2014}
Jones, A.~P., Ysard, N., Koehler, M., {et~al.} 2014, Faraday Discuss., 168,
  313, arXiv: 1411.5877

\bibitem[{Juvela(2019)}]{juvela_soc_2019}
Juvela, M. 2019, Astronomy and Astrophysics, 622, A79

\bibitem[{Juvela {et~al.}(2018{\natexlab{a}})Juvela, Guillet, Liu, Ristorcelli,
  Pelkonen, Alina, Bronfman, Eden, Kim, Koch, Kwon, Lee, Malinen, Micelotta,
  Montillaud, Rawlings, Sanhueza, Soam, Traficante, Ysard, \&
  Zhang}]{juvela_dust_2018}
Juvela, M., Guillet, V., Liu, T., {et~al.} 2018{\natexlab{a}}, Astronomy and
  Astrophysics, 620, A26

\bibitem[{Juvela {et~al.}(2018{\natexlab{b}})Juvela, Malinen, Montillaud,
  Pelkonen, Ristorcelli, \& Tóth}]{juvela_galactic_2018}
Juvela, M., Malinen, J., Montillaud, J., {et~al.} 2018{\natexlab{b}}, Astronomy
  and Astrophysics, 614, A83

\bibitem[{Juvela {et~al.}(2019)Juvela, Padoan, Ristorcelli, \&
  Pelkonen}]{juvela_synthetic_2019}
Juvela, M., Padoan, P., Ristorcelli, I., \& Pelkonen, V.-M. 2019, Astronomy \&
  Astrophysics, 629, A63

\bibitem[{Juvela {et~al.}(2015)Juvela, Ristorcelli, Marshall, Montillaud,
  Pelkonen, Ysard, McGehee, Paladini, Pagani, Malinen, Rivera-Ingraham,
  Lefèvre, Tóth, Montier, Bernard, \& Martin}]{juvela_galactic_2015}
Juvela, M., Ristorcelli, I., Marshall, D.~J., {et~al.} 2015, Astronomy and
  Astrophysics, 584, A93

\bibitem[{Juvela {et~al.}(2011)Juvela, Ristorcelli, Pelkonen, Marshall,
  Montier, Bernard, Paladini, Lunttila, Abergel, André, Dickinson, Dupac,
  Malinen, Martin, McGehee, Pagani, Ysard, \& Zavagno}]{juvela_galactic_2011}
Juvela, M., Ristorcelli, I., Pelkonen, V.-M., {et~al.} 2011, Astronomy \&
  Astrophysics, 527, A111

\bibitem[{Kim {et~al.}(1994)Kim, Martin, \& Hendry}]{kim_size_1994}
Kim, S.-H., Martin, P.~G., \& Hendry, P.~D. 1994, The Astrophysical Journal,
  422, 164

\bibitem[{Kramer {et~al.}(1996)Kramer, Stutzki, \&
  Winnewisser}]{kramer_structure_1996}
Kramer, C., Stutzki, J., \& Winnewisser, G. 1996, Astronomy and Astrophysics,
  307, 915

\bibitem[{Köhler {et~al.}(2014)Köhler, Jones, \& Ysard}]{kohler_hidden_2014}
Köhler, M., Jones, A., \& Ysard, N. 2014, Astronomy and Astrophysics, 565, L9

\bibitem[{Köhler {et~al.}(2015)Köhler, Ysard, \& Jones}]{kohler_dust_2015}
Köhler, M., Ysard, N., \& Jones, A.~P. 2015, Astronomy and Astrophysics, 579,
  A15

\bibitem[{Lada {et~al.}(1991)Lada, Bally, \& Stark}]{lada_unbiased_1991}
Lada, E.~A., Bally, J., \& Stark, A.~A. 1991, The Astrophysical Journal, 368,
  432

\bibitem[{Laureijs {et~al.}(1991)Laureijs, Clark, \&
  Prusti}]{laureijs_iras_1991}
Laureijs, R.~J., Clark, F.~O., \& Prusti, T. 1991, The Astrophysical Journal,
  372, 185

\bibitem[{Le~Gal {et~al.}(2017)Le~Gal, Herbst, Dufour, Gratier, Ruaud, Vidal,
  \& Wakelam}]{le_gal_new_2017}
Le~Gal, R., Herbst, E., Dufour, G., {et~al.} 2017, Astronomy and Astrophysics,
  605, A88

\bibitem[{Le~Petit {et~al.}(2006)Le~Petit, Nehmé, Le~Bourlot, \&
  Roueff}]{le_petit_model_2006}
Le~Petit, F., Nehmé, C., Le~Bourlot, J., \& Roueff, E. 2006, The Astrophysical
  Journal Supplement Series, 164, 506

\bibitem[{Li \& Draine(2001)}]{li_ultrasmall_2001}
Li, A. \& Draine, B.~T. 2001, The Astrophysical Journal Letters, 550, L213

\bibitem[{Li \& Greenberg(1997)}]{li_unified_1997}
Li, A. \& Greenberg, J.~M. 1997, Astronomy and Astrophysics, 323, 566

\bibitem[{Martin {et~al.}(2012)Martin, Roy, Bontemps, Miville-Deschênes, Ade,
  Bock, Chapin, Devlin, Dicker, Griffin, Gundersen, Halpern, Hargrave, Hughes,
  Klein, Marsden, Mauskopf, Netterfield, Olmi, Patanchon, Rex, Scott, Semisch,
  Truch, Tucker, Tucker, Viero, \& Wiebe}]{martin_evidence_2012}
Martin, P.~G., Roy, A., Bontemps, S., {et~al.} 2012, The Astrophysical Journal,
  751, 28

\bibitem[{Mathis {et~al.}(1977)Mathis, Rumpl, \& Nordsieck}]{mathis_size_1977}
Mathis, J.~S., Rumpl, W., \& Nordsieck, K.~H. 1977, The Astrophysical Journal,
  217, 425

\bibitem[{Mathis \& Whiffen(1989)}]{mathis_composite_1989}
Mathis, J.~S. \& Whiffen, G. 1989, The Astrophysical Journal, 341, 808

\bibitem[{Milman {et~al.}(1973)Milman, Knapp, Knapp, \&
  Wilson}]{milman_co_1973}
Milman, A.~S., Knapp, G.~R., Knapp, S.~L., \& Wilson, W.~J. 1973, in , 332

\bibitem[{Neckel \& Sarcander(1985)}]{neckel_spectroscopic_1985}
Neckel, T. \& Sarcander, M. 1985, Astronomy and Astrophysics, 147, L1

\bibitem[{Ohashi {et~al.}(2013)Ohashi, Kitamura, \&
  Akashi}]{ohashi_mapping_2013}
Ohashi, S., Kitamura, Y., \& Akashi, T. 2013, in , 345

\bibitem[{Ormel {et~al.}(2011)Ormel, Min, Tielens, Dominik, \&
  Paszun}]{ormel_dust_2011}
Ormel, C.~W., Min, M., Tielens, A. G. G.~M., Dominik, C., \& Paszun, D. 2011,
  Astronomy and Astrophysics, 532, A43

\bibitem[{Ossenkopf \& Henning(1994)}]{ossenkopf_dust_1994}
Ossenkopf, V. \& Henning, T. 1994, Astronomy and Astrophysics, 291, 943

\bibitem[{Pety {et~al.}(2007)Pety, Goicoechea, Hily-Blant, Gerin, \&
  Teyssier}]{pety_deuterium_2007}
Pety, J., Goicoechea, J.~R., Hily-Blant, P., Gerin, M., \& Teyssier, D. 2007,
  Astronomy and Astrophysics, 464, L41

\bibitem[{Pety {et~al.}(2012)Pety, Gratier, Guzmán, Roueff, Gerin, Goicoechea,
  Bardeau, Sievers, Le~Petit, Le~Bourlot, Belloche, \&
  Talbi}]{pety_iram-30_2012}
Pety, J., Gratier, P., Guzmán, V., {et~al.} 2012, Astronomy and Astrophysics,
  548, A68

\bibitem[{Pety {et~al.}(2005)Pety, Teyssier, Fossé, Gerin, Roueff, Abergel,
  Habart, \& Cernicharo}]{pety_are_2005}
Pety, J., Teyssier, D., Fossé, D., {et~al.} 2005, Astronomy and Astrophysics,
  435, 885

\bibitem[{Philipp {et~al.}(2006)Philipp, Lis, Güsten, Kasemann, Klein, \&
  Phillips}]{philipp_submillimeter_2006}
Philipp, S.~D., Lis, D.~C., Güsten, R., {et~al.} 2006, Astronomy and
  Astrophysics, 454, 213

\bibitem[{Pilleri {et~al.}(2012)Pilleri, Montillaud, Berné, \&
  Joblin}]{pilleri_evaporating_2012}
Pilleri, P., Montillaud, J., Berné, O., \& Joblin, C. 2012, Astronomy and
  Astrophysics, 542, A69

\bibitem[{Pilleri {et~al.}(2015)Pilleri, Reisenfeld, Zurbuchen, Lepri, Shearer,
  Gilbert, von Steiger, \& Wiens}]{pilleri_variations_2015}
Pilleri, P., Reisenfeld, D.~B., Zurbuchen, T.~H., {et~al.} 2015, The
  Astrophysical Journal, 812, 1

\bibitem[{{Planck Collaboration} {et~al.}(2011){Planck Collaboration}, Abergel,
  Ade, Aghanim, Arnaud, Ashdown, Aumont, Baccigalupi, Balbi, Banday, Barreiro,
  Bartlett, Battaner, Benabed, Benoît, Bernard, Bersanelli, Bhatia, Blagrave,
  Bock, Bonaldi, Bond, Borrill, Bouchet, Boulanger, Bucher, Burigana, Cabella,
  Cantalupo, Cardoso, Catalano, Cayón, Challinor, Chamballu, Chiang, Chiang,
  Christensen, Clements, Colombi, Couchot, Coulais, Crill, Cuttaia, Danese,
  Davies, Davis, de~Bernardis, de~Gasperis, de~Rosa, de~Zotti, Delabrouille,
  Delouis, Désert, Dickinson, Donzelli, Doré, Dörl, Douspis, Dupac,
  Efstathiou, Enßlin, Eriksen, Finelli, Forni, Frailis, Franceschi, Galeotta,
  Ganga, Giard, Giardino, Giraud-Héraud, González-Nuevo, Górski, Gratton,
  Gregorio, Gruppuso, Hansen, Harrison, Helou, Henrot-Versillé, Herranz,
  Hildebrandt, Hivon, Hobson, Holmes, Hovest, Hoyland, Huffenberger, Jaffe,
  Joncas, Jones, Jones, Juvela, Keihänen, Keskitalo, Kisner, Kneissl, Knox,
  Kurki-Suonio, Lagache, Lamarre, Lasenby, Laureijs, Lawrence, Leach, Leonardi,
  Leroy, Linden-Vørnle, Lockman, López-Caniego, Lubin, Macías-Pérez,
  MacTavish, Maffei, Maino, Mandolesi, Mann, Maris, Marshall, Martin,
  Martínez-González, Masi, Matarrese, Matthai, Mazzotta, McGehee, Meinhold,
  Melchiorri, Mendes, Mennella, Miville-Deschênes, Moneti, Montier, Morgante,
  Mortlock, Munshi, Murphy, Naselsky, Nati, Natoli, Netterfield,
  Nørgaard-Nielsen, Noviello, Novikov, Novikov, O'Dwyer, Osborne, Pajot,
  Paladini, Pasian, Patanchon, Perdereau, Perotto, Perrotta, Piacentini, Piat,
  Pinheiro~Gonçalves, Plaszczynski, Pointecouteau, Polenta, Ponthieu,
  Poutanen, Prézeau, Prunet, Puget, Rachen, Reach, Reinecke, Renault,
  Ricciardi, Riller, Ristorcelli, Rocha, Rosset, Rowan-Robinson,
  Rubiño-Martín, Rusholme, Sandri, Santos, Savini, Scott, Seiffert, Shellard,
  Smoot, Starck, Stivoli, Stolyarov, Stompor, Sudiwala, Sygnet, Tauber,
  Terenzi, Toffolatti, Tomasi, Torre, Tristram, Tuovinen, Umana, Valenziano,
  Vielva, Villa, Vittorio, Wade, Wandelt, Wilkinson, Yvon, Zacchei, \&
  Zonca}]{planck_collaboration_planck_2011-1}
{Planck Collaboration}, Abergel, A., Ade, P. A.~R., {et~al.} 2011, Astronomy
  and Astrophysics, 536, A24

\bibitem[{Pound {et~al.}(2003)Pound, Reipurth, \& Bally}]{pound_looking_2003}
Pound, M.~W., Reipurth, B., \& Bally, J. 2003, The Astronomical Journal, 125,
  2108

\bibitem[{Reach {et~al.}(2005)Reach, Megeath, Cohen, Hora, Carey, Surace,
  Willner, Barmby, Wilson, Glaccum, Lowrance, Marengo, \&
  Fazio}]{reach_absolute_2005}
Reach, W., Megeath, S., Cohen, M., {et~al.} 2005, Publications of the
  Astronomical Society of the Pacific, 117, 978

\bibitem[{Roy {et~al.}(2013)Roy, Martin, Polychroni, Bontemps, Abergel, André,
  Arzoumanian, Di~Francesco, Hill, Konyves, Nguyen-Luong, Pezzuto, Schneider,
  Testi, \& White}]{roy_changes_2013}
Roy, A., Martin, P.~G., Polychroni, D., {et~al.} 2013, The Astrophysical
  Journal, 763, 55

\bibitem[{Sandell {et~al.}(1986)Sandell, Reipurth, Menten, Walmsley, \&
  Ungerechts}]{sandell_young_1986}
Sandell, G., Reipurth, B., Menten, C., Walmsley, M., \& Ungerechts, H. 1986, in
  , 295

\bibitem[{Schaerer \& de~Koter(1997)}]{schaerer_combined_1997}
Schaerer, D. \& de~Koter, A. 1997, Astronomy and Astrophysics, 322, 598

\bibitem[{Siebenmorgen \& Kruegel(1992)}]{siebenmorgen_dust_1992}
Siebenmorgen, R. \& Kruegel, E. 1992, Astronomy and Astrophysics, 259, 614

\bibitem[{Smith(1984)}]{smith_optical_1984}
Smith, F.~W. 1984, Journal of Applied Physics, 55, 764

\bibitem[{Stansberry {et~al.}(2007)Stansberry, Gordon, Bhattacharya,
  Engelbracht, Rieke, Marleau, Fadda, Frayer, Noriega‐Crespo, Wachter, Young,
  Müller, Kelly, Blaylock, Henderson, Neugebauer, Beeman, \&
  Haller}]{stansberry_absolute_2007}
Stansberry, J., Gordon, K., Bhattacharya, B., {et~al.} 2007, Publications of
  the Astronomical Society of the Pacific, 119, 1038

\bibitem[{Stark \& Bally(1982)}]{stark_co_1982}
Stark, A.~A. \& Bally, J. 1982, in , 329--333

\bibitem[{Stepnik {et~al.}(2003)Stepnik, Abergel, Bernard, Boulanger,
  Cambrésy, Giard, Jones, Lagache, Lamarre, Meny, Pajot, Le~Peintre,
  Ristorcelli, Serra, \& Torre}]{stepnik_evolution_2003}
Stepnik, B., Abergel, A., Bernard, J.-P., {et~al.} 2003, Astronomy and
  Astrophysics, 398, 551

\bibitem[{Swinyard {et~al.}(2010)Swinyard, Ade, Baluteau, Aussel, Barlow,
  Bendo, Benielli, Bock, Brisbin, Conley, Conversi, Dowell, Dowell, Ferlet,
  Fulton, Glenn, Glauser, Griffin, Griffin, Guest, Imhof, Isaak, Jones, King,
  Leeks, Levenson, Lim, Lu, Makiwa, Naylor, Nguyen, Oliver, Panuzzo,
  Papageorgiou, Pearson, Pohlen, Polehampton, Pouliquen, Rigopoulou, Ronayette,
  Roussel, Rykala, Savini, Schulz, Schwartz, Shupe, Sibthorpe, Sidher, Smith,
  Spencer, Trichas, Triou, Valtchanov, Wesson, Woodcraft, Xu, Zemcov, \&
  Zhang}]{swinyard_-flight_2010}
Swinyard, B.~M., Ade, P., Baluteau, J.-P., {et~al.} 2010, Astronomy and
  Astrophysics, 518, L4

\bibitem[{Teyssier {et~al.}(2004)Teyssier, Fossé, Gerin, Pety, Abergel, \&
  Roueff}]{teyssier_carbon_2004}
Teyssier, D., Fossé, D., Gerin, M., {et~al.} 2004, Astronomy and Astrophysics,
  417, 135

\bibitem[{Van De~Putte {et~al.}(2019)Van De~Putte, Gordon, Roman-Duval,
  Williams, Baes, Tchernyshyov, Lawton, \& Arab}]{van_de_putte_evidence_2019}
Van De~Putte, D., Gordon, K.~D., Roman-Duval, J., {et~al.} 2019, The
  Astrophysical Journal, 888, 22

\bibitem[{Wakelam {et~al.}(2017)Wakelam, Bron, Cazaux, Dulieu, Gry, Guillard,
  Habart, Hornekær, Morisset, Nyman, Pirronello, Price, Valdivia, Vidali, \&
  Watanabe}]{wakelam_h$_2$_2017}
Wakelam, V., Bron, E., Cazaux, S., {et~al.} 2017, arXiv:1711.10568 [astro-ph],
  arXiv: 1711.10568

\bibitem[{Warren \& Hesser(1977)}]{warren_photometric_1977}
Warren, Jr., W.~H. \& Hesser, J.~E. 1977, The Astrophysical Journal Supplement
  Series, 34, 115

\bibitem[{Weingartner \& Draine(2001{\natexlab{a}})}]{weingartner_dust_2001}
Weingartner, J.~C. \& Draine, B.~T. 2001{\natexlab{a}}, The Astrophysical
  Journal, 548, 296

\bibitem[{Weingartner \&
  Draine(2001{\natexlab{b}})}]{weingartner_photoelectric_2001}
Weingartner, J.~C. \& Draine, B.~T. 2001{\natexlab{b}}, The Astrophysical
  Journal Supplement Series, 134, 263

\bibitem[{Ysard {et~al.}(2013)Ysard, Abergel, Ristorcelli, Juvela, Pagani,
  Könyves, Spencer, White, \& Zavagno}]{ysard_variation_2013}
Ysard, N., Abergel, A., Ristorcelli, I., {et~al.} 2013, Astronomy and
  Astrophysics, 559, A133

\bibitem[{Ysard {et~al.}(2016)Ysard, Köhler, Jones, Dartois, Godard, \&
  Gavilan}]{ysard_mantle_2016}
Ysard, N., Köhler, M., Jones, A., {et~al.} 2016, Astronomy and Astrophysics,
  588, A44

\bibitem[{Ysard {et~al.}(2015)Ysard, Köhler, Jones, Miville-Deschênes,
  Abergel, \& Fanciullo}]{ysard_dust_2015}
Ysard, N., Köhler, M., Jones, A., {et~al.} 2015, Astronomy and Astrophysics,
  577, A110

\bibitem[{Zhou {et~al.}(1993)Zhou, Jaffe, Howe, Geis, Herrmann, Madden,
  Poglitsch, \& Stacey}]{zhou_[c_1993}
Zhou, S., Jaffe, D.~T., Howe, J.~E., {et~al.} 1993, The Astrophysical Journal,
  419, 190

\bibitem[{Zubko {et~al.}(2004)Zubko, Dwek, \& Arendt}]{zubko_interstellar_2004}
Zubko, V., Dwek, E., \& Arendt, R.~G. 2004, The Astrophysical Journal
  Supplement Series, 152, 211

\end{thebibliography}

\begin{appendix}
\section{Size distribution}
\label{appendix:size_distribution}
Size distributions of dust in  THEMIS follow either 
a power-law with an exponential cut-off, defined as follow : 
\begin{equation}
    \frac{\mathrm{d}n}{\mathrm{d}a} \propto \left\{
                            \begin{array}{ll}
                                a^{\alpha}  & \qquad \mathrm{if} \quad a < a_{\mathrm{t}}   \\
                                a^{\alpha}  \times \exp\left(-\left(\frac{a-a_{\mathrm{t}}}{a_{\mathrm{c}}}\right)^{3}\right) & \qquad \mathrm{if} \quad a \geq a_{\mathrm{t}}
                            \end{array}
                           \right.
\end{equation}

\noindent or a log-normal law, defined as follow :
\begin{equation}
    \frac{\mathrm{d}n}{\mathrm{d}a} \propto \frac{1}{a} \times \exp\left(-\left(\frac{\log(a/a_{0})}{\sigma}\right)^{2}\right)
\end{equation}
\noindent where all the parameters for each dust distribution are
listed in Table\,\ref{tab:parameters_size_distribution}.

\begin{table}
\caption{\label{tab:parameters_size_distribution} Size distribution parameters
for each dust population (see Appendix\,\ref{appendix:size_distribution} for
the equations). p-law is a power-law with an exponential tail and log-n is 
a log-normal distribution.}
\centering
\begin{tabular}{lccccccc}
    \hline\hline 
     Name & size & $\alpha$ & $a_{\mathrm{min}}$ & $a_{\mathrm{max}}$ & 
     $a_{\mathrm{c}}$ & $a_{\mathrm{t}}$ & $a_{0}$  \\
     \hline 
     & &  & [nm] & [nm] & [nm] & [nm] & [nm] \\ 
     \hline 
     \multicolumn{8}{c}{Core Mantle grains (CM)} \\
     \cm & p-law & 5 & 0.4 & 4900 & 10 & 50 & - \\
     \CM & log-n & - & 0.5 & 4900 & - & - & 7 \\
     \py & log-n & - & 1 & 4900 & - & - & 8 \\
     \hline
     \multicolumn{8}{c}{Aggregated Mantle Mantle grains (AMM)} \\
     AMM & log-n & - & 47.9 & 700 & - & - & 479 \\
     \hline
     \multicolumn{8}{c}{Aggregated Mantle Mantle Ice grains (AMMI)} \\
     AMMI & log-n & - & 91.2 & 700 & - & - & 610 \\
     \hline 
\end{tabular}
\end{table}

\section{The Horsehead seen with \spitzer~and \herschel}
\label{appendix:HH_obs}

\begin{figure*}[h]
\centering
	\includegraphics[width=0.33\textwidth, trim={0 0cm 0cm 0cm},clip]{1_HH_0.pdf}\hfill
	\includegraphics[width=0.33\textwidth, trim={0 0cm 0cm 0cm},clip]{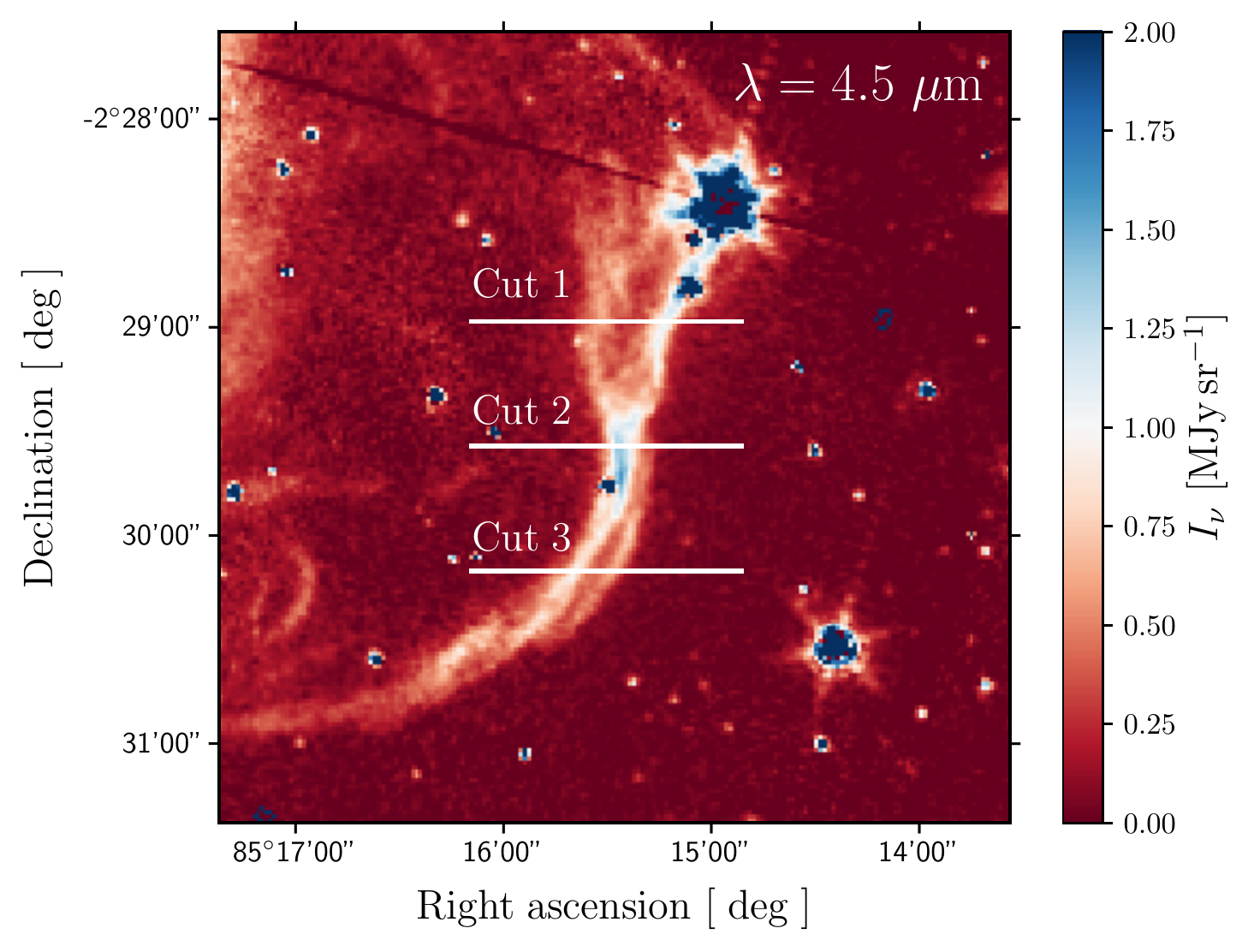}\hfill
	\includegraphics[width=0.33\textwidth, trim={0 0cm 0cm 0cm},clip]{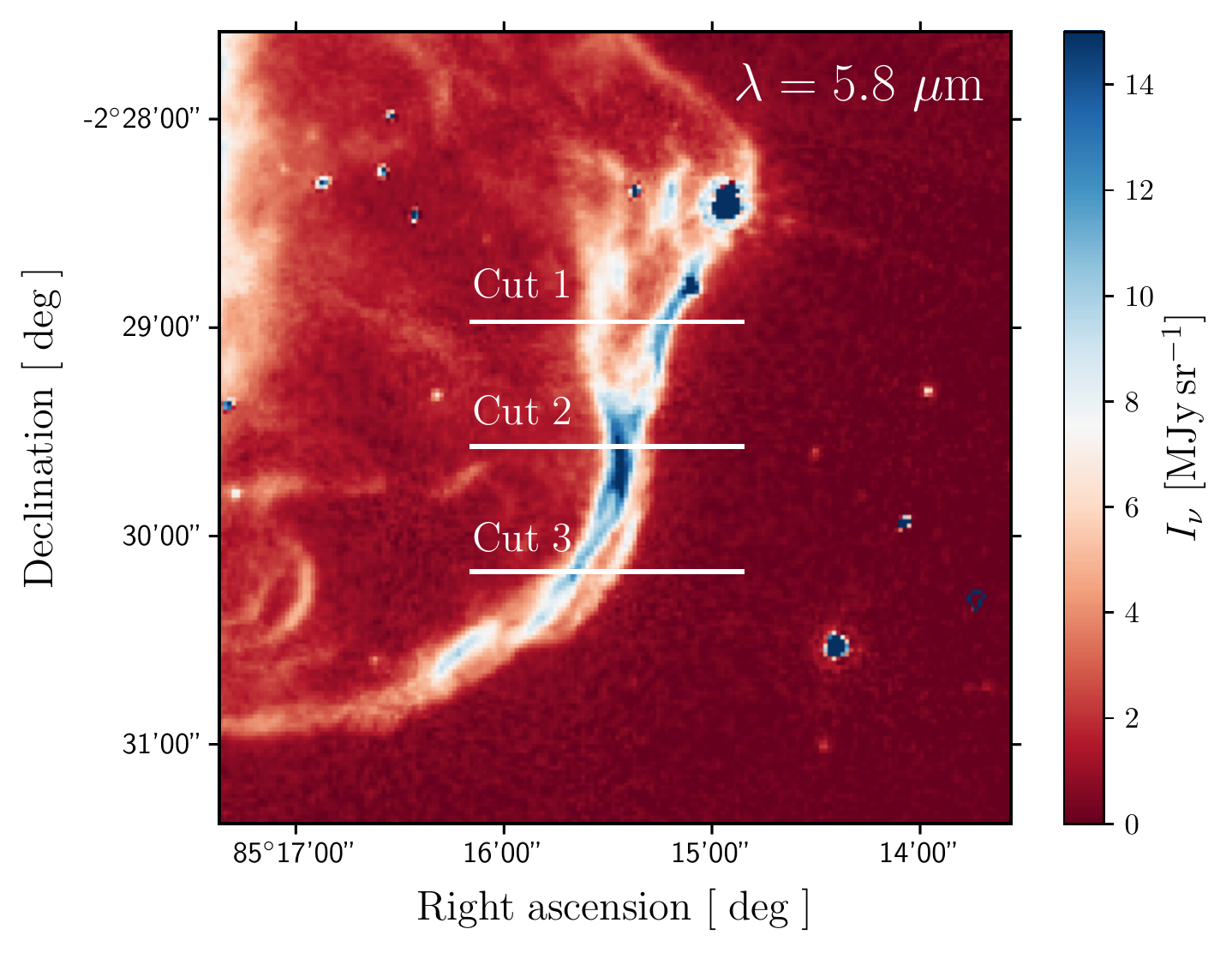}\\
	\includegraphics[width=0.33\textwidth, trim={0 0cm 0cm 0cm},clip]{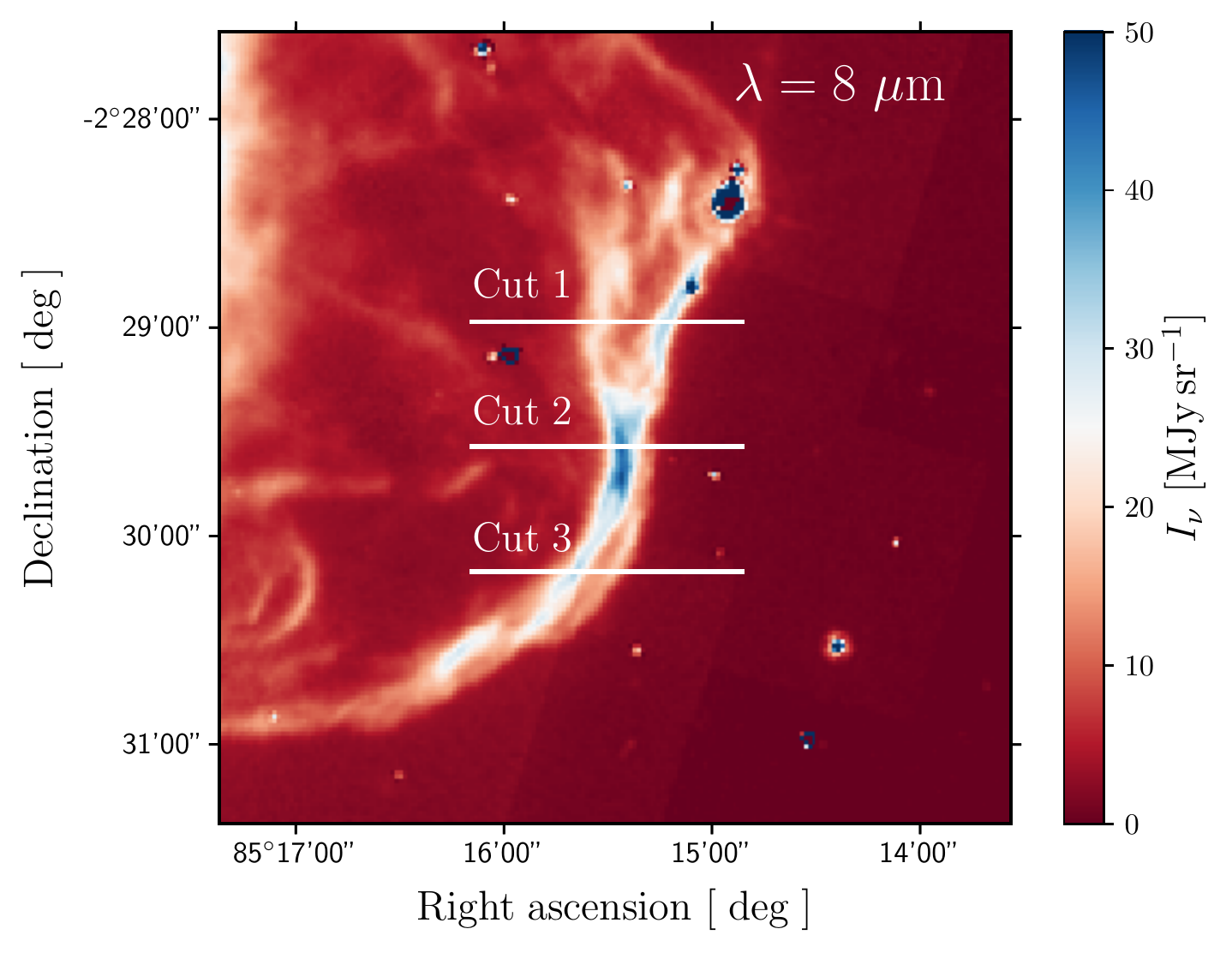}\hfill
	\includegraphics[width=0.33\textwidth, trim={0 0cm 0cm 0cm},clip]{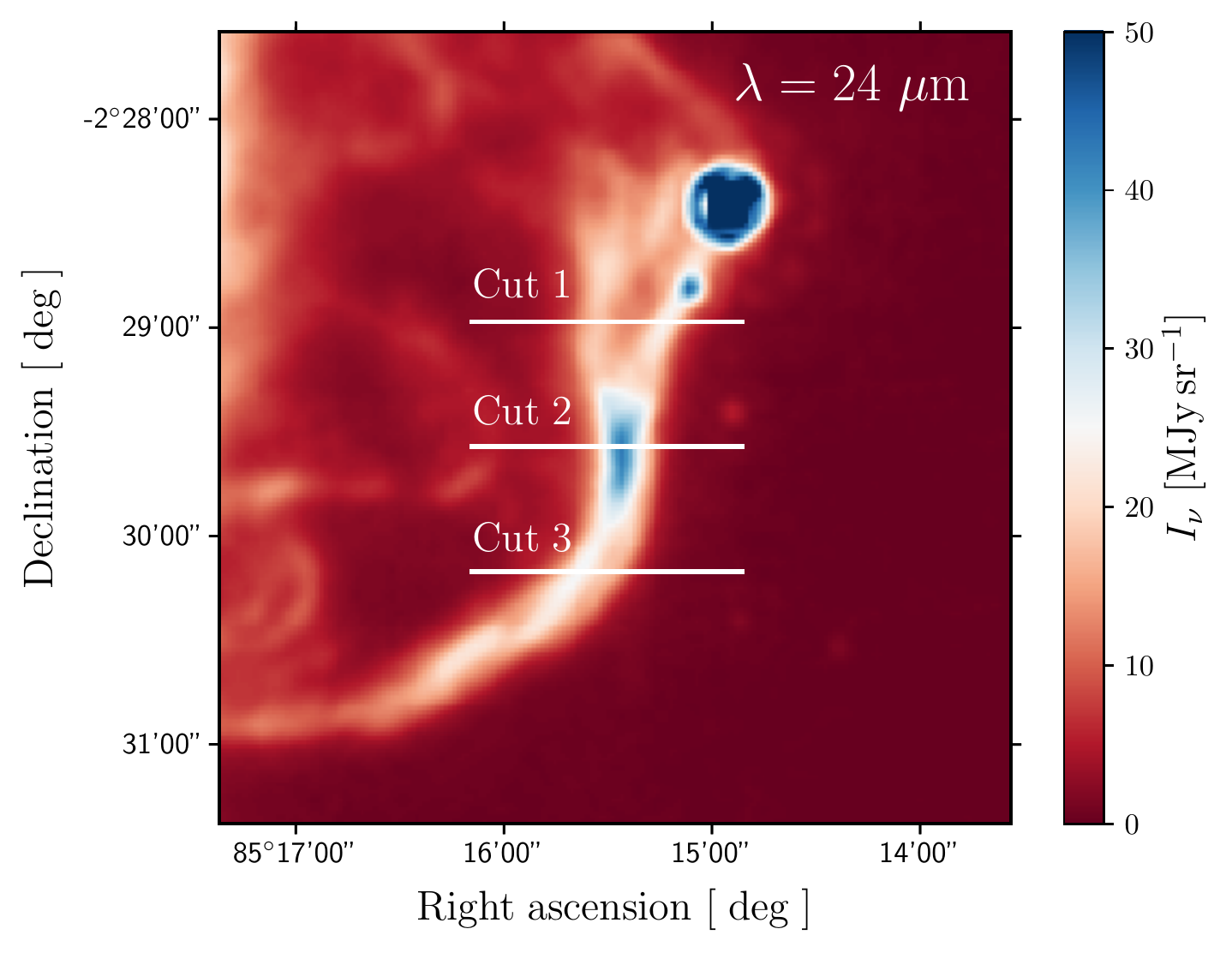}\hfill
	\includegraphics[width=0.33\textwidth, trim={0 0cm 0cm 0cm},clip]{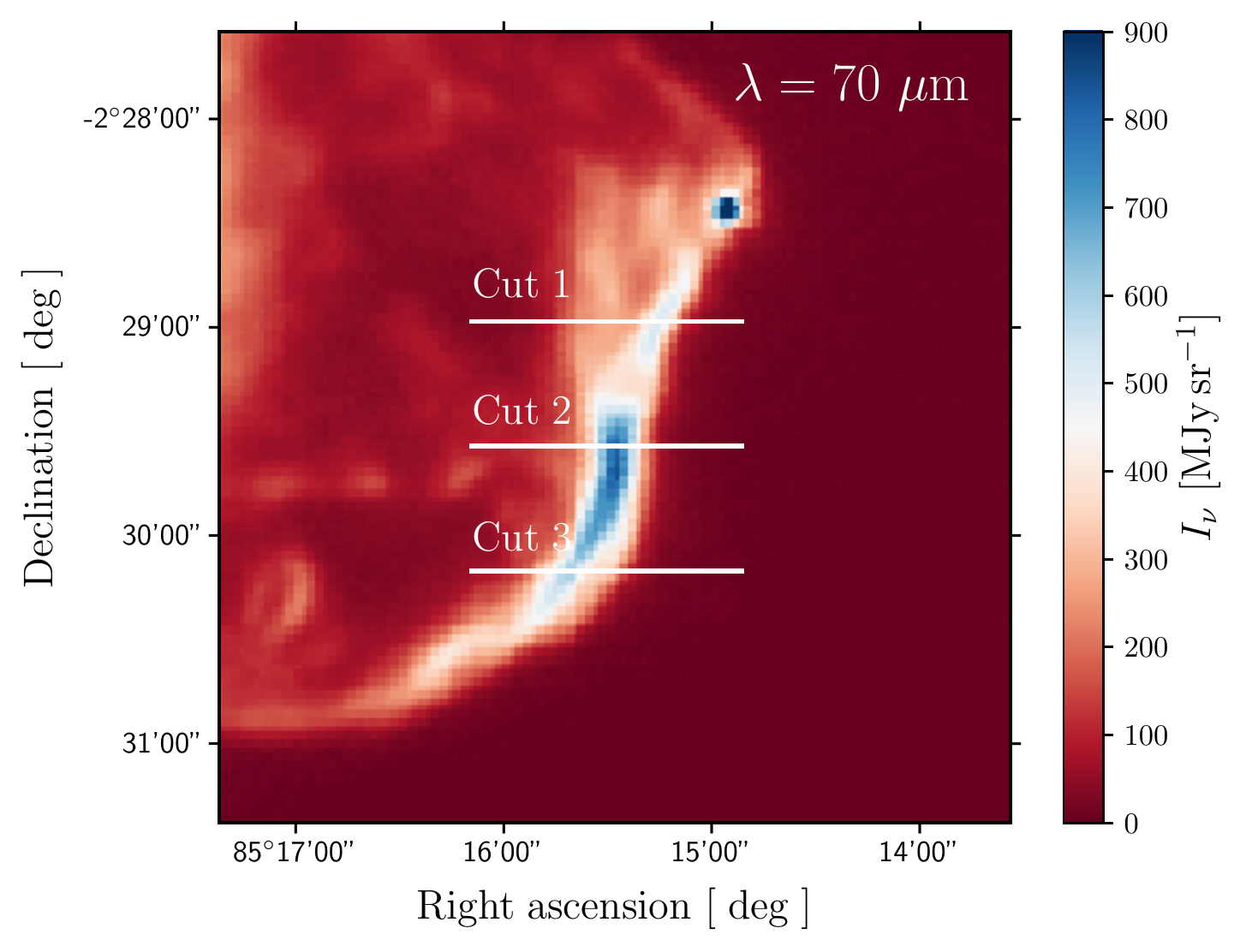}\\
	\includegraphics[width=0.33\textwidth, trim={0 0cm 0cm 0cm},clip]{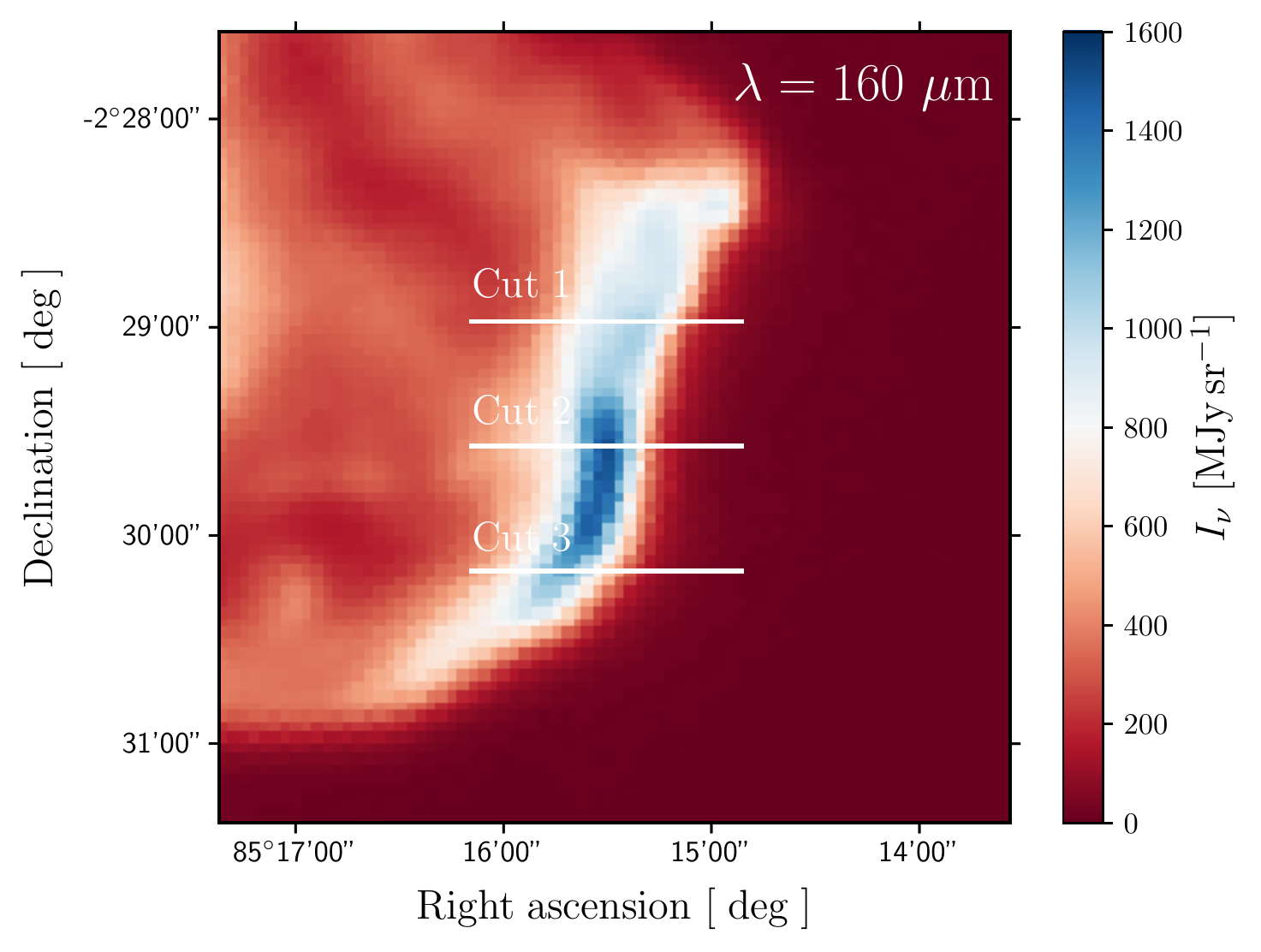}\hfill
	\includegraphics[width=0.33\textwidth, trim={0 0cm 0cm 0cm},clip]{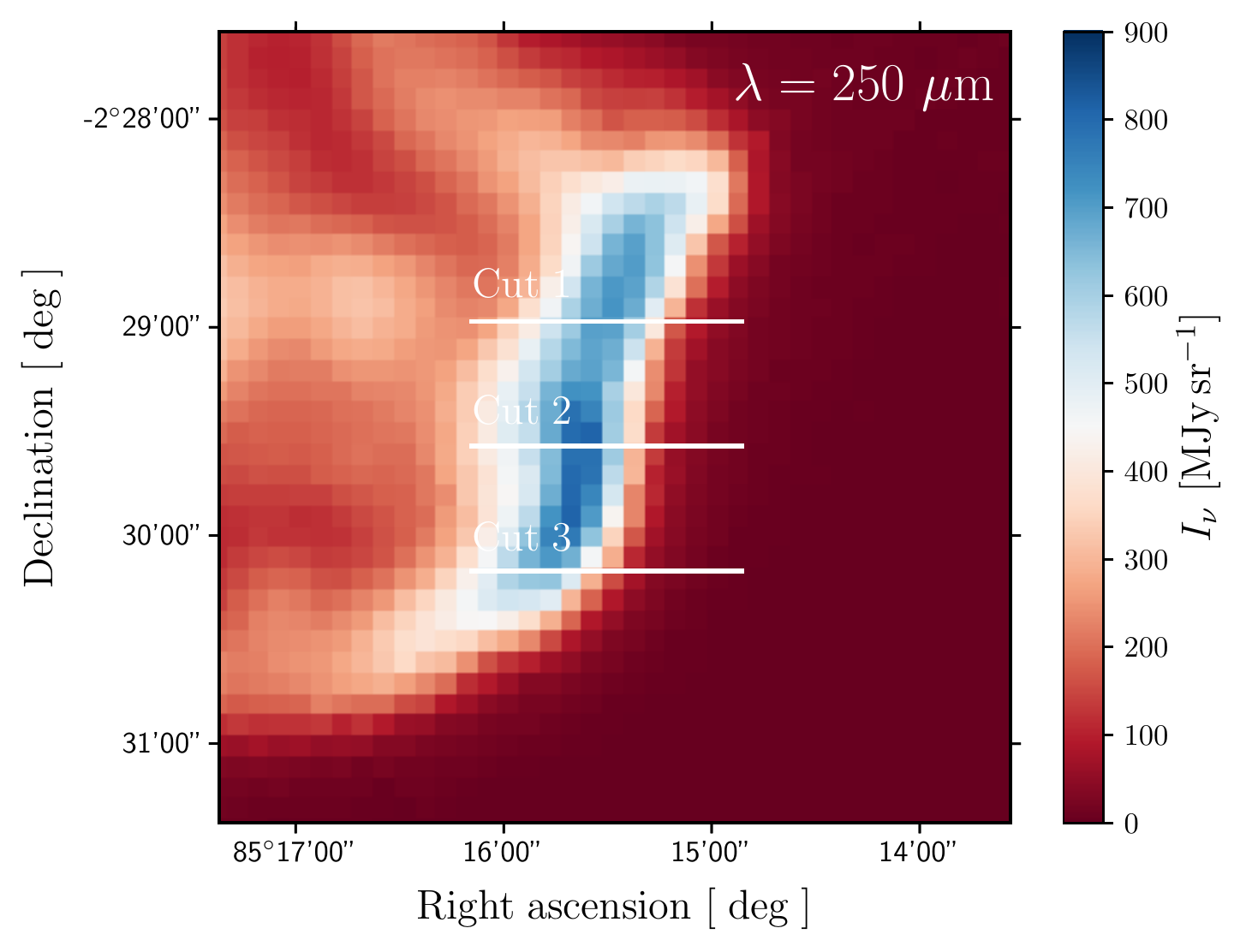}\hfill
	\includegraphics[width=0.33\textwidth, trim={0 0cm 0cm 0cm},clip]{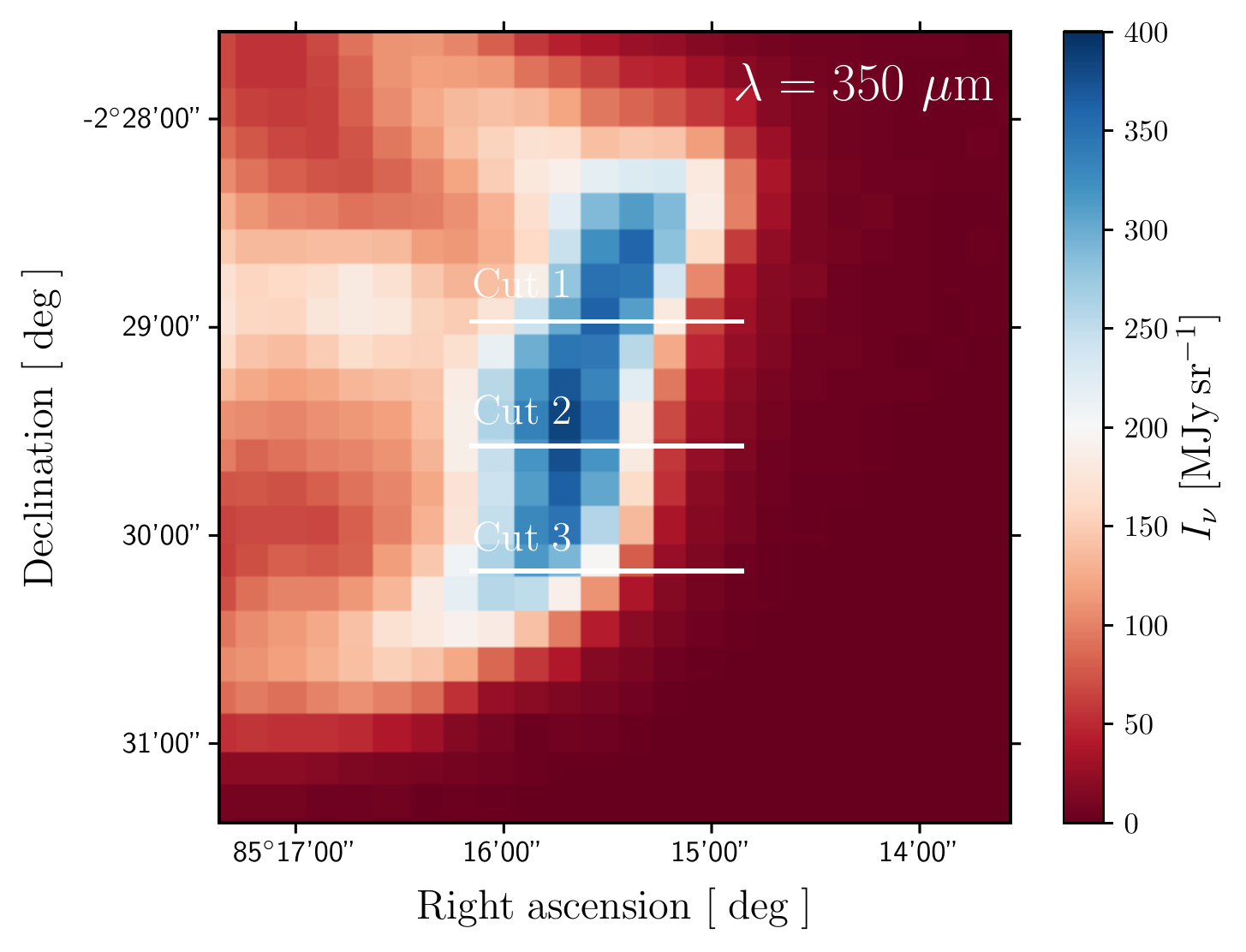}\\
	\includegraphics[width=0.33\textwidth, trim={0 0cm 0cm 0cm},clip]{1_HH_9.pdf}
    \caption{The Horsehead seen in the 10 photometric bands. The 3 white solid
    lines correspond to the 3 cuts we use in our study.}
    \label{fig:HH_tot_1}
\end{figure*}

\end{appendix}

%
%

\end{document}